\documentclass{aa} % [referee]
\usepackage{amsmath}
\usepackage{times,graphics,epsfig,color,geometry}
\usepackage[colorlinks=true,allcolors=blue]{hyperref}
\usepackage{natbib}
\usepackage[caption=false]{subfig} % [caption=false]
\usepackage{multirow}
\usepackage{parskip}
\usepackage{array}
\usepackage{stfloats}
\usepackage{hhline}
\usepackage{xfrac}
\usepackage{multirow}
\DeclareMathOperator{\sech}{sech}

\begin{document}

\title{Towards a fully consistent Milky Way disk model V. \\ 
The disk model for 4\,--\,14 kpc} 
% Age-metallicity relation from  RC stars 
% based on Gaia and APOGEE 

\author{K. Sysoliatina\inst{1} and A. Just\inst{1}
}

%\offprints{K. Sysoliatina}
\mail{Sysoliatina@uni-heidelberg.de}

\institute{$^{1}$Astronomisches Rechen-Institut, Zentrum f\"{u}r Astronomie der Universit\"{a}t Heidelberg, 
M\"{o}nchhofstr. 12--14, 69120 Heidelberg, \mbox{Germany}}

\date{Printed: \today}

\abstract 
%Context
{The semi-analytic Just-Jahrei{\ss} (JJ) model of the Galactic disk 
is a flexible tool for stellar population synthesis with a fine age resolution 
of \mbox{25 Myr}. The model has recently been 
calibrated in the solar neighbourhood against the \textit{Gaia} DR2 stars. 
We have identified two star-formation bursts 
within the last $\sim$4 Gyr of the local star-formation rate (SFR) history.}
%Aims
{In this work we present a generalised version of the JJ model that incorporates 
our findings for the solar neighbourhood and is applicable to a wide range of 
galactocentric distances, 4 kpc $\lesssim R \lesssim $ 14 kpc.}
%Methods
{The JJ model includes the four flattened and two spheroidal components of the Milky Way, 
describing it as an axisymmetric system. 
The thin and thick disks, as well as atomic and molecular gas layers, are assumed to have
exponential radial surface density profiles. Spherical stellar halo and dark matter  
in the form of a cored isothermal sphere are also added to the model. 
The overall thickness of the thin disk is assumed to be constant at all radii, 
though model realisations with a flaring disk can also be tested.
The adopted radial variation in the thin-disk SFR reflects the inside-out disk 
growth scenario. Motivated by our findings for the solar neighbourhood, we allow a smooth
power-law SFR continuum to be modified by an arbitrary number of Gaussian peaks. 
Additionally, the vertical kinematics of the stellar populations associated with 
these episodes of star-formation excess is allowed to differ from the kinematics prescribed by 
the age-velocity dispersion relation for the thin-disk populations of the same age.}
%Results 
{We present a public code of the JJ model complemented by the three sets of isochrones
generated by the stellar tracks and isochrones with the PAdova and TRieste Stellar 
Evolution Code (PARSEC), the Modules and Experiments in Stellar Astrophysics 
(MESA) Isochrones and Stellar Tracks (MIST), and a Bag of Stellar Tracks and Isochrones
(BaSTI). Assuming a plausible set of parameters,
we take the first step towards calibrating \mbox{the JJ model} at non-solar radii.  
Using metallicity distributions of the red clump giants from the Apache Point Observatory Galactic Evolution Experiment (APOGEE), we constrain 
the radial variation of \mbox{the JJ-model} age-metallicity relation and propose 
a new analytic form for the age-metallicity relation function.}
%Conclusions 
{The generalised JJ model is a publicly available tool for studying different 
stellar populations across the Milky Way disk. With its fine age resolution and flexibility, 
it can be particularly useful for reconstructing the thin-disk SFR, 
as a variety of different SFR shapes can be constructed within its framework.}


\authorrunning{K. Sysoliatina and A. Just}
\titlerunning{The global Milky Way disk model}

\maketitle 

%\parskip=0.pt
%%%%%%%%%%%%%%%%%%%%%%%%%%%%%%%%%%%%%%%%%%%%%%%%%%%%%%%%%%%%%%%%%%%%%
%%%%%%%%%%%%%%%%%%%%%%%%%%%%%%%%%%%%%%%%%%%%%%%%%%%%%%%%%%%%%%%%%%%%%
\defcitealias{just10}{Paper~I}
\defcitealias{just11}{Paper~II}
\defcitealias{rybizki15}{Paper~III}
\defcitealias{sysoliatina21}{Paper~IV}

\section{Introduction}\label{sect:intro}

During the last few decades, extensive astrometric, photometric, and spectroscopic 
data on the Milky Way (MW\footnote{See \mbox{Table \ref{tab:acronyms} for the list of
abbreviations.}}) stellar populations have been collected. As a result, it became
possible to build complex and realistic MW models and constrain their parameters by
comparing their predictions to the observed stellar distributions, motions, chemical
abundances, and ages. And though no exhaustive MW model exists at the moment, 
significant progress has been achieved towards understanding the MW 
evolutionary processes.

Relatively simple semi-analytic MW models, such as the TRI-dimensional modeL 
of thE GALaxy (TRILEGAL\footnote{\url{http://stev.oapd.inaf.it/cgi-bin/trilegal}}), 
\mbox{a priori} prescribe some observationally motivated spatial density 
distributions for the Galactic components. With the help of a stellar evolution library 
and additional assumptions about the MW star-formation rate 
(SFR) history, initial mass function (IMF), and age-metallicity relation (AMR), these densities 
are then transformed into star counts within some observational field \citep{girardi05,girardi16}. 
%These kinds of models are not self-consistent in a sense that no global MW gravitational potential is derived from the assumed density profiles 
%and vice versa -- the density profiles' shapes are not brought into consistency with 
%the potential they produce. However, even following this simple approach, 
Even with this simple approach it is possible to obtain
realistic star count predictions for the volumes where the model has been trained; 
as a result, the optimised model parameters can tell us a lot about the evolution 
of the MW stellar components \citep{pieres20,daltio21}. 

%To be self-consistent, and thus more realistic, a MW model needs to be based 
%on the principles of stellar dynamics. A big group of models investigates 
%the Galactic potential via stellar distribution functions (DF) of angle-action variables,
%usually under the assumption of dynamical equilibrium \citep{binney12,binney14,bovy13}. 
%The DF can be then extended to other physical parameters of the stars, 
%such as metallicity, and thus, a chemodynamical MW model can be built \citep{sanders15}. 
%At the next step, non-axisymmetric features can be added, such as bar and spiral arms \citep{monari16,kazwini20}. 

One of the most comprehensive semi-analytic MW models to date 
is the Besan\c con Galaxy model (BGM\footnote{\url{https://model.obs-besancon.fr}};
\citealp{robin3,robin12,robin17,czekaj14,bienayme18}). This dynamically self-consistent
model describes gaseous, stellar, and dark matter (DM) MW content in terms of sets 
of sub-components characterised by the different kinematics, chemistry, ages, 
and densities. The BGM also includes non-axisymmetric features, such as the bar,  
spiral arms, disk flare, and warp \citep{amores17}, and accounts for extinction. 
The model has many applications in the field of Galactic studies -- for example, \citet{mor19}
optimised its predictions using the second data release (DR2) of the \textit{Gaia} 
mission \citep{gaia16a,gaia18} and find evidence of an increase 
in star-formation (SF) activity in the last \mbox{$\sim$2--3 Gyr}  
of the MW disk history. However, the results of that work do not allow the authors to draw 
a firm conclusion about the SFR shape because of the limited age resolution of the model: 
there are only seven sub-populations representing the thin disk in the BGM, and this 
may be not enough to decipher with certainty the evolutionary complexity of the MW disk. 

During the last decade, we have been developing a semi-analytic model of the MW disk 
with a time resolution as fine as \mbox{25 Myr} 
(\citealp{just10}, hereafter \citetalias{just10}). 
So rather than having seven sub-populations as in the BGM, 
the Just-Jahrei{\ss} (JJ) model builds thin disk from 520 mono-age sub-populations. 
This allows the disk structure and evolution to be studied in great detail, which 
also allows theoretical modelling to keep up with the ever-growing amount and quality 
of data accumulated for the MW stellar populations.

The JJ model is based on an iterative solving of the 
Poisson equation, and thus reconstructs a self-consistent pair of the overall vertical 
gravitational potential and density profile. The thin disk is modelled in the presence 
of gas, thick disk, stellar halo, and DM components. The disk evolution is governed 
by four input functions: the SFR, IMF, age-velocity dispersion relation (AVR), 
and AMR. So far, \mbox{the JJ model} has been  
tested many times and calibrated against different data in the solar neighbourhood. 
In \citet{just11} (hereafter \citetalias{just11}) we used star counts towards 
the north Galactic pole from the seventh data release (DR7) of the Sloan Digital 
Sky Survey (SDSS; \citealp{abazajian09}) to optimise the SFR and thick-disk parameters. 
\citet{rybizki15} (hereafter \citetalias{rybizki15}) improved the IMF with the help of 
data provided by the HIgh Precision PARallax COllecting Satellite 
(\textsc{Hipparcos}; \citealp{vanleeuwen07}) complemented by an updated version of 
the Catalogue of Nearby Stars (CNS 4; \citealp{jahreiss97}). % (CNS5). 
In \citet{sysoliatina18} we demonstrated a good model-to-data consistency 
(with a $\sim$7.3-\% discrepancy in terms of star counts) in the local 1-kpc height 
cylinder using stars from the Radial Velocity Experiment survey 
(RAVE; \citealp{steinmetz06,kunder17}) 
and the Tycho-Gaia Astrometric Solution catalogue (TGAS; \citealp{michalik15,lindegren16}).
Most recently, we used \textit{Gaia} DR2 stars within \mbox{600 pc} of the Sun  
and simultaneously optimised 22 model parameters within the Bayesian framework
(\citealp{sysoliatina21}, hereafter \citetalias{sysoliatina21}). 
We also calibrated the AMR against the local red clump (RC) sample from 
the Apache Point Observatory Galactic Evolution Experiment survey 
(APOGEE; \citealp{eisenstein11,bovy14,majewski17}). 
We have identified two recent SF burst episodes 
in the thin-disk evolutionary history, which happened \mbox{$\sim0.5$ Gyr} and 
\mbox{$\sim3$ Gyr} ago and left behind dynamically heated fossil populations. 

In this paper we generalise our local JJ model for the range of galactocentric distances 
\mbox{4 kpc $\lesssim R \lesssim$ 14 kpc} by building the disk out of a set of independent 
radial annuli. We assume radially exponential gaseous, thin and thick disks, 
a spherical stellar halo, and a cored isothermal DM halo. We fix the disk thickness 
by assuming a constant-thickness thick disk and a constant-thickness or flaring thin disk  
and allow radial variations in the input functions 
SFR, AVR, and AMR, which correspond to the inside-out disk growth scenario. 
We present a publicly available tool, python package \texttt{jjmodel}, 
which can be used to perform stellar population synthesis 
within \mbox{the JJ-model} framework. 
Following \citetalias{sysoliatina21}, we constrain the thin- and thick-disk AMR
using the generalised \mbox{JJ model} with the radially extended APOGEE RC sample. 
We then propose a new analytic form for the thin-disk AMR, which is able to comprise 
the variation in its shape across the studied distances, \mbox{4 kpc $< R <$ 14 kpc}. 
Thus, we add another dimension to the fine JJ-model resolution: 
the numerous MW mono-age sub-populations can be now synthesised over a fine $R$-$z$ grid.

This paper has the following structure. In \mbox{Section \ref{sect:model}} 
we review \mbox{the JJ-model} components, explain how different ingredients 
of the local \mbox{JJ model} are now extended to other galactocentric distances, 
and summarise the overall scheme of mock sample modelling. 
\mbox{Section \ref{sect:model_predictions}} contains a description of 
a test model and its predictions for the MW disk properties.  
In \mbox{Section \ref{sect:amr}} we reconstruct the radially dependent AMR using 
APOGEE RC data. Then, we discuss the obtained results for AMR, as well as 
the generalised \mbox{JJ-model} predictions and their possible applications, 
in \mbox{Section \ref{sect:discussion}}. Finally, we summarise the main aspects 
of this work in \mbox{Section \ref{sect:conclusions}}. 

\begin{table}[t!]
    \centering
    \caption{List of abbreviations used in this paper.}
    \begin{tabular}{lc} \hhline{==} \\[-0.2cm]
    AMR & Age-metallicity relation \\
    AVR & Age-velocity dispersion relation \\
    CADF & Cumulative age distribution function \\
    CMDF & Cumulative metallicity distribution function \\ 
    IMF & Initial mass function \\ 
    MD & Metallicity distribution \\ 
    MW & Milky Way \\
    RC & Red clump \\ 
    SF & Star formation \\ 
    SFR & Star-formation rate \vspace{0.1cm}  \\  \hhline{--}
    \end{tabular}
    \label{tab:acronyms}
\end{table}

\section{Generalised JJ model}\label{sect:model}

In this section we introduce our key assumptions about the structure and 
properties of the Galaxy (\mbox{Section \ref{sect:assumptions}}), recall 
stellar dynamic equations used for calculation of the Galactic potential 
(\mbox{Section \ref{sect:vert_prof}}), and review our treatment of the MW components 
in the local \mbox{JJ model} to generalise it for 
the whole MW disk (\mbox{Sections \ref{sect:thindisk}-\ref{sect:bulge}}). 
We also briefly remind our approach to converting the modelled densities into 
mock samples (\mbox{Section \ref{sect:sps}}) and present an overall scheme of building 
the MW disk within \mbox{the JJ-model} framework (\mbox{Section \ref{sect:scheme}}). 

\subsection{Basic assumptions and limitations}\label{sect:assumptions}

Before going into details, we briefly summarise the main assumptions on which  
\mbox{the JJ model} is based. We also mention limitations imposed by these 
assumptions on the model applicability (more in \mbox{Section \ref{sect:vert_prof}}).
%\vspace{-0.25em}

% Version for A&A where lists are not allowed outside conclusion...
\ifx
First, the MW disk rests in a steady state. This statement is in harmony 
with the observational data, which show no evidence of fast evolutionary 
processes happening in our Galaxy at the present day (no major merger, 
no activity in the nucleus). All transformations of the MW populations' 
distribution functions occur slowly, as a result of collective interactions 
of the MW components (the so-called secular evolution). Thus, for our 
purpose of describing the MW at the present-day time snapshot, 
it is safe to treat the system as a steady one. 

Second, the MW disk is axisymmetric and plane-symmetric 
(no bar, spiral arms, or warp are included). 
This implies that \mbox{the JJ model} cannot be used at small radii, 
\mbox{$R \lesssim$ 4 kpc}, where this assumption breaks because 
of a presence of the bar. Also, for comparison to the data, unless the model 
is used to reveal 
the underlying deviations from the smooth density distribution, \mbox{the JJ-model} 
predictions need to be averaged over large enough volumes to smear out small-scale density perturbations, such as overdensities in the spiral arms. 
Due to the absence of the warp, the predicted vertical structure can be 
reliable up to moderate radii, \mbox{$R \lesssim 14$ kpc}. 

Third, the system is flattened (radial and vertical dynamics can be decoupled). 
Within the range of considered galactocentric distances, 
\mbox{4 kpc $ < R < $ 14 kpc}, this holds up to \mbox{$|z| \approx$ 2 kpc} 
(see \mbox{Section \ref{sect:vert_prof}} for details).

Fourth, the MW stellar, gaseous, and DM components are represented by an individual 
or multiple isothermal sub-populations. Some of these sub-populations are assumed to be 
globally isothermal (stellar and DM halo), while others are isothermal only locally 
at fixed $R$ (thin disk, thick disk, molecular and atomic gas). 

Fifth, all sub-populations are in equilibrium with the total gravitational potential 
(Poisson and Jeans equations can be applied). This may not hold for very young populations,
which did not have enough time to reach the equilibrium. 
Thus, the model cannot be applied to the youngest populations (\mbox{O- and} B-type stars). 
\fi 

% I leave list for the arxiv version. I really want to have our assumptions 
% collected in one list. 
\begin{itemize}
\setlength\itemsep{0.5em} 
    \item The MW disk rests in a steady state. This statement is in harmony 
    with the observational data, which show no evidence of fast evolutionary 
    processes happening in our Galaxy at the present day (no major merger, 
    no activity in the nucleus). All transformations of the MW populations' 
    distribution functions occur slowly, as a result of collective interactions 
    of the MW components (the so-called secular evolution). Thus, for our 
    purpose of describing the MW at the present-day time snapshot, 
    it is safe to treat the system as a steady one. 
    
    \item The MW disk is axisymmetric and plane-symmetric 
    (no bar, spiral arms, or warp are included). 
    This implies that \mbox{the JJ model} cannot be used at small radii, 
    \mbox{$R \lesssim$ 4 kpc}, where this assumption breaks because 
    of a presence of the bar. Also, for comparison to the data, unless the model 
    is used to reveal 
    the underlying deviations from the smooth density distribution, \mbox{the JJ-model} 
    predictions need to be averaged over large enough volumes to smear out small-scale density perturbations, such as overdensities in the spiral arms. 
    Due to the absence of the warp, the predicted vertical structure can be 
    reliable up to moderate radii, \mbox{$R \lesssim 14$ kpc}. 
    
    \item The system is flattened (radial and vertical dynamics can be decoupled). 
    Within the range of considered galactocentric distances, 
    \mbox{4 kpc $ < R < $ 14 kpc}, this holds up to \mbox{$|z| \approx$ 2 kpc} 
    (see \mbox{Section \ref{sect:vert_prof}} for details).
    
    \item The MW stellar, gaseous, and DM components are represented by an individual 
    or multiple isothermal sub-populations. Some of these sub-populations are assumed to be 
    globally isothermal (stellar and DM halo), while others are isothermal only locally 
    at fixed $R$ (thin disk, thick disk, molecular and atomic gas). 
    
    \item All sub-populations are in equilibrium with the total gravitational potential 
    (Poisson and Jeans equations can be applied). This may not hold for very young populations,
    which did not have enough time to reach the equilibrium. 
    Thus, the model cannot be applied to the youngest populations (\mbox{O- and} B-type stars). 
\end{itemize}

\subsection{Vertical profiles}\label{sect:vert_prof}

At each galactocentric distance, the overall density in \mbox{the JJ model} 
is prescribed as a sum of $i$ different MW components: stellar 
(thin disk, thick disk, stellar halo), gaseous (molecular and atomic gas), and DM\footnote{Throughout this paper, subscripts `d', `t', `sh', `dh', `H$_2$', and `HI' 
stand for the thin disk, thick disk, stellar halo, DM halo, molecular and atomic gas, respectively. To avoid confusion, we also apply the following index notation.
Index $i$ is reserved for the MW components; mono-age isothermal sub-populations of 
the MW components are associated with index $j$; index $k$ corresponds 
to extra thin-disk SFR peaks (\mbox{Section \ref{sect:thindisk_sfr}}); finally, 
stellar assemblies (\mbox{Section \ref{sect:sps}}) are used with index $l$.}:
\begin{equation}
\rho(\Phi) = \sum_{i} \rho_\mathrm{i} = 
\rho_\mathrm{d} + \rho_\mathrm{t} + \rho_\mathrm{sh} + 
\rho_\mathrm{H_2} + \rho_\mathrm{HI} + \rho_\mathrm{dh}.
\label{eq:rhoPhi_tot}
\end{equation}
Here $\Phi$ is the total gravitational potential produced by this matter. 
For a single isothermal sub-population with the vertical velocity dispersion 
$\sigma_\mathrm{z,j}$, its density is linked to the total gravitational potential
according to the Jeans equation:
\begin{equation}
 \sigma_\mathrm{z,j}^2 \frac{\partial \rho_\mathrm{j}}{\partial z} + 
 \rho_\mathrm{j} \frac{\partial \Phi}{\partial z} = 0.  
    \label{eq:JE_iso}
\end{equation}
Here we dropped a time-derivative term assuming a steady state. We also ignored 
two cross-terms containing $\sigma_\mathrm{Rz}$, essentially implying  
that the radial and vertical motions in the Galactic disk are decoupled.
It can be shown that the ratio of these cross-terms 
to the rest of the terms in the Jeans equation is of the order of 
$z/R \, \cdot \, h_\mathrm{j}/R_\mathrm{j}$, where $h_\mathrm{j}$ and $R_\mathrm{j}$ 
are the sub-population's scale height and scale length, 
respectively \citep{bahcall84,bienayme87}. In a flattened MW-like system, 
$h_\mathrm{j} \ll R_\mathrm{j}$, and therefore, within the range of radii 
\mbox{4 kpc $ < R < $ 14 kpc} this ratio is small up to \mbox{$|z|\approx 2$ kpc}. 

The solution of \mbox{Eq. (\ref{eq:JE_iso})} implies that the vertical density 
fall-off of the isothermal sub-population is an exponential function 
of the total gravitational potential: 
\begin{equation}
    \rho_\mathrm{j} = \rho_\mathrm{j0} \exp{\left( -\Phi/\sigma_\mathrm{z,j}^2 \right)}. 
    \label{eq:rhoz_expPhi}
\end{equation}
Here potential is zero in the Galactic plane, 
and $\rho_\mathrm{j0}$ is the sub-population's midplane density  
defined via its surface density $\Sigma_\mathrm{j}$ and scale height $h_\mathrm{j}$:
\begin{equation}
\rho_\mathrm{j0} = \frac{\Sigma_\mathrm{j}}{2 h_\mathrm{j}}. 
    \label{eq:rho_j0}
\end{equation}

As a next step, we assume that all MW components in \mbox{Eq. (\ref{eq:rhoPhi_tot})}
can be represented by a single or multiple mono-age isothermal sub-populations, 
whose vertical density profiles follow \mbox{Eq. (\ref{eq:rhoz_expPhi})}. 
As a result, \mbox{Eq. (\ref{eq:rhoPhi_tot})} can be turned into a specific function 
of the total potential with the vertical velocity dispersions and midplane densities of 
sub-populations as model parameters. Then, we can substitute this function into 
the Poisson equation and find a self-consistent potential-density pair. 

For an axisymmetric system, the Poisson equation in cylindrical coordinates can be 
written as 
\begin{equation}
    \frac{\partial^2 \Phi(R,z)}{\partial z^2} = 
    4 \pi G \rho(R,z) + \frac{1}{R}\frac{\partial}{\partial R} \left( R  F_\mathrm{R} \right),
    \label{eq:PE}
\end{equation}
where $G$ is the gravitational constant and 
$F_\mathrm{R} = - \partial \Phi(R,z) / \partial R$ is the radial force. 
It is common to assume that \mbox{in flattened} systems
the second term on the right-hand side (the radial term) is not important 
in comparison to the first one (the density term), and thus, can be neglected \citep{binney_tremaine}. Indeed, as a zero-order approximation, $R F_\mathrm{R}$ 
can be replaced by its value in the plane, namely, by the squared circular velocity 
$\upsilon_\mathrm{c}^2(R)$. As the MW rotation curve is nearly flat in a wide range 
of galactocentric distances, the radial term is vanishingly small and can be dropped. 
For the rotation curve reconstruction in 
\citet{sysoliatina18}, we developed a more accurate treatment 
of the radial force that also includes dependence on $z$ (see Eq. (4.19) in 
\citealp{sysoliatina18_thesis}). Thus, we can estimate a ratio of the radial 
to density terms, in the framework of our MW mass model used in \citet{sysoliatina18}. 
We find that in the range of galactocentric distances 
\mbox{4 kpc $ < R < $ 14 kpc}, this ratio is $\sim$0.01 in the Galactic plane 
and may become important only at \mbox{$|z| \approx 2 $ kpc} in the inner disk, 
\mbox{$R \lesssim $ 5 kpc}, where it reaches \mbox{$\sim$0.2--0.5}. But even in this case, 
the radial term can still be neglected as it remains smaller 
than the density term uncertainty. The latter is dominated by the uncertainty of 
a poorly known DM density, which typically reaches tens of percents \citep{read14}. 
% \citet{bovy12}

With these considerations taken into account, we neglect the radial term 
in \mbox{Eq. (\ref{eq:PE})} that significantly simplifies our problem. 
The Poisson equation without the radial term states that the vertical gravitational 
potential at some radius is produced only by matter at this radius, that is to say, 
the radial and vertical structure of the disk can be treated separately. 
This means that our approach of building the MW disk out of a set of independent 
radial annuli is justified. 
At each radius, the solution of the simplified Poisson equation reads as 
\begin{equation}
z(\Phi) = \int_0^{\Phi} \left( 8 \pi G \int_0^{\Phi^\prime}   
\rho (\Phi^{\prime\prime}) \mathrm{d}\Phi^{\prime\prime} \right)^{-1/2} \mathrm{d}\Phi^\prime. 
\label{eq:zPhi}
\end{equation}
Here $\rho(\Phi)$ is the total density 
at this radius given by \mbox{Eqs. (\ref{eq:rhoPhi_tot})} with substitution from 
\mbox{Eqs. (\ref{eq:rhoz_expPhi}) and (\ref{eq:rho_j0})}. 

In the following subsections, we describe in detail how each MW component 
is modelled over the considered range of galactocentric radii.

\subsection{Thin disk}\label{sect:thindisk}

There are four analytic functions in the \mbox{JJ model}, 
which reflect main aspects of the MW thin-disk evolution and describe its 
present-day state: IMF, SFR, AVR, and AMR. The thin disk has a present-day age of 
\mbox{$t_\mathrm{p} = 13$ Gyr} and consists of a set of locally isothermal 
sub-populations covering the whole age range with a fine resolution 
of $t_\mathrm{r} = 25$ Myr. 
This results into 520 thin-disk sub-populations, each characterised by 
its present-day mean age, metallicity, vertical velocity dispersion, and 
surface density. Below we explain how each of these quantities are prescribed
\mbox{(Sections \ref{sect:thindisk_imf}--\ref{sect:thindisk_avr})}, 
and how, together with the thick-disk and halo sub-populations, they are eventually 
converted into star counts \mbox{(Section \ref{sect:sps})}.  
 
\subsubsection{IMF}\label{sect:thindisk_imf}

For the IMF, we adopt a four-slope broken power law. Its up-to-date 
parameters are listed in \citetalias{sysoliatina21} (\mbox{Table 4} and 
\mbox{Section 4.6.2}) where we optimised them in consistency with the kinematic 
and SFR parameters using the local \textit{Gaia} DR2 sample. 
This default \mbox{JJ-model} IMF is also applied during stellar population synthesis 
of the thick disk and halo (Section \ref{sect:sps}). In the \texttt{jjmodel} 
package, the user can both change the IMF parameters and customise 
the model by defining another IMF law. 

It is worth mentioning here that there is different observational evidence
indicating that IMF is more top-heavy in the environments with low metallicity 
and high density \citep{bartko10,banerjee12,kalari18}, 
and thus, also in the early Universe. 
On the other side, massive early-type galaxies seem to have a bottom-heavy IMF, 
and the same is reported for the MW low-metallicity stars in \citet{hallakoun21}.
A possibility for IMF varying in space and time is discussed in the literature 
\citep{kroupa20} and tested with various methods -- for example, in the cosmological
simulations of the \mbox{MW-like} galaxies \citep{guszejnov17} or with semi-analytic 
models of SF in pre-stellar cores of different metallicities 
\citep{tanaka17,tanaka18,tanaka19}. 
However, most of the observed IMF variations are moderate and can be fitted 
into the universal IMF paradigm \citep{bastian10,offner14}. 
In our model we chose to stick to the conventional universal IMF 
for all disk sub-populations at all $R$ and $z$. We do this for two reasons: 
(1) spatially and time-dependent IMF remains debatable, and (2) the regions 
we model have never belonged to the environments with extreme physical conditions, 
such as the Galactic centre, where the most significant deviations 
of the canonical IMF are observed. 

\subsubsection{AMR}\label{sect:thindisk_amr}

The AMR function describes in a simple way the chemical evolution of the disk 
by linking the metallicity and age of its stellar sub-populations. 
In the previous \mbox{JJ-model}
versions, we used for the thin-disk AMR a four-parameter logarithmic law
(\mbox{Eqs. (21) and (22)} \mbox{in \citetalias{sysoliatina21}}). 
It describes a relatively quick 
chemical enrichment during the first \mbox{$\sim$0.5--1 Gyr} of the disk evolution 
with a subsequent quasi-linear increase of metallicity with age in 
the solar neighbourhood. 

There are two options for introducing the thin-disk AMR 
in the \mbox{\texttt{jjmodel} code}. More sophisticated and default AMR 
is described by a seven-parameter $\tanh$-law, whose parameters are constants, 
linear or broken linear laws of $R$ (Section \ref{sect:amr_fit}). 
By construction, this AMR is consistent 
with the observed metallicity distribution (MD) of the APOGEE RC stars 
across the disk, the reconstruction process of this AMR is described in 
\mbox{Section \ref{sect:amr}}. 

The second and simpler way to prescribe the thin-disk AMR is 
analogous to how we extend the local thin-disk SFR to other radii
\mbox{(Section \ref{sect:thindisk_sfr})}: parameters of the four-parameter 
logarithmic AMR function with known values at $R_\odot$ (\mbox{Table 3} 
in \citetalias{sysoliatina21}) are assumed to vary with radius according to 
power laws with predefined slopes.

\subsubsection{SFR}\label{sect:thindisk_sfr}

Starting from \citetalias{just10}, the default thin-disk SFR was a power-law 
function with a peak at \mbox{$\sim$10 Gyr} ago and a monotonous decline  
to the present day. In \citetalias{sysoliatina21}, we found that the thin-disk SFR
prescribed in such a way is inconsistent with the observed \mbox{\textit{Gaia} DR2}
star counts of young populations in the local volume. The data-to-model
consistency can be significantly improved by adding to the declining SFR continuum 
two Gaussian peaks centred at ages of \mbox{$\sim$0.5 Gyr} and \mbox{$\sim$3 Gyr}. 
This result is consistent with \citet{mor19} who find a single SF burst centred 
at the age of \mbox{2--3 Gyr}. Also, \citet{ruiz-lara20} report  
three SF enhancements \mbox{5.7 Gyr}, \mbox{1.9 Gyr}, and \mbox{1 Gyr} ago,  
which were presumably triggered by recurrent passages of the Sagittarius dwarf galaxy.
\citet{sahlholdt22} studied disk subgiants and dwarfs from the Third Data Release of the 
Galactic Archaeology with HERMES (GALAH DR3; \citealp{buder21}) and 
\textit{Gaia} Early Data Release 3 
(EDR3; \citealp{gaia21}). The authors have also identified several enhancements in the disk SFR 
with the most recent SF peak at ages \mbox{2--4 Gyr}, in agreement with the above mentioned studies.
Motivated by these findings, we now define the thin-disk SFR as a standard smooth 
continuum that can be modified by an arbitrary number of overlaying Gaussian peaks, 
which allows the construction and testing of SFR functions of complicated shapes. 

At a fixed radius, we define the thin-disk SFR as follows: 
\begin{equation}
\begin{aligned}
SFR_\mathrm{d} &= \left< SFR_\mathrm{d} \right> \cdot sfr_\mathrm{d}(t) = \\ 
&= \left< SFR_\mathrm{d} \right> \cdot (sfr_\mathrm{dc}(t) + sfr_\mathrm{dp}(t))
\end{aligned}
\label{eq:SFRd}
\end{equation}
It is a product of a normalisation 
constant and a time-dependent function; the latter consists of two terms that are related 
to the continuum and additional peaks (denoted by `dc' and `dp', respectively).
The function is normalised to the overall present-day thin-disk surface density $\Sigma_\mathrm{d}$:
\begin{equation}
\left< SFR_\mathrm{d} \right> = \frac{\Sigma_\mathrm{d}}   
{\int_0^{t_\mathrm{p}} sfr_\mathrm{d}(t) g_\mathrm{d}(t) dt}. 
\label{eq:sfrdn}    
\end{equation}
A mass loss function $g_\mathrm{d}(t)$ in \mbox{Eq. (\ref{eq:sfrdn})} 
accounts for the cumulative effect of stellar evolution 
and is needed to convert the present-day stellar mass into the mass at the time 
of formation. In particular, $g_\mathrm{d}(t)$ gives a fraction of 
a sub-population's mass survived to the present day in the form of stars 
(including remnants). This fraction depends on the IMF and the sub-population's 
metallicity (i.e., AMR; for $g_\mathrm{d}(t)$ expression see \mbox{Eq. (9)} 
in \citetalias{sysoliatina21}). We calculate these fractions with the default 
\mbox{JJ-model} IMF over the grid of metallicities and ages using 
the chemical evolution code \textit{Chempy}\footnote{\url{https://github.com/jan-rybizki/Chempy}} 
\citep{rybizki17}. For each realisation of \mbox{the JJ model}, the mass loss 
function is then constructed in consistency with the chosen AMR by interpolation 
over the tabulated values. As the mass loss function depends on the IMF very weakly,
and an impact of the mass loss function itself on the modelled quantities is 
of a higher order than of SFR and AVR, we do not provide a tabulated 
$g_\mathrm{d}(t)$ for IMF laws other than our standard four-slope broken power law. 

The continuum part of the thin-disk SFR is given by \mbox{Eq. (10)} 
in \citetalias{sysoliatina21}, and for the sake of completeness we also 
reproduce it here:
\begin{equation}
sfr_\mathrm{dc}(t) = \frac{(t^2-t_\mathrm{d1}^2)^\zeta}{(t + t_\mathrm{d2})^\eta}.
\label{eq:sfrdc}    
\end{equation}
In this equation, parameters $t_\mathrm{d1}$, $t_\mathrm{d2}$, $\zeta$, 
and $\eta$ define the function's peak position and amplitude, as well as 
the steepness of the declining part. Following \citetalias{sysoliatina21},
we define each extra SFR peak as a Gaussian weighted by the value 
of the SFR continuum at the peak's mean age and normalised to the 
continuum maximum value:
\begin{align}
\label{eq:sfrdp}
sfr_\mathrm{dp}(t) = \sum_k^{N_\mathrm{p}} sfr_\mathrm{dc}(t_\mathrm{p} &-\tau_\mathrm{p,k}) 
\times \\ \nonumber        
 &\times \frac{\Sigma_\mathrm{p,k}}{\Sigma_\mathrm{max}} 
 \exp{\left( - \frac{(t_\mathrm{p} - \tau_\mathrm{p,k} - t)^2}{2 (d\tau_\mathrm{p,k})^2}  \right)},
\end{align}
\vspace{-10pt}
\begin{equation}
\text{where} \quad \Sigma_\mathrm{max} = \frac{\Sigma_\mathrm{d}}  
{\int_0^{t_\mathrm{p}} sfr_\mathrm{dc}(t) g_\mathrm{d}(t) dt} 
sfr_\mathrm{dc}(t_\mathrm{max}). \nonumber
\end{equation}
Here $\Sigma_\mathrm{p}$ is related to a peak's amplitude at a given radius, 
and parameters $\tau_\mathrm{p}$ and $d\tau_\mathrm{p}$ give the mean age 
and the age dispersion of the peak, respectively. Galactic time $t_\mathrm{max}$
corresponds to the peak of the SFR continuum $\Sigma_\mathrm{max}$ 
(\mbox{Eq. (11)}\footnote{This equation in \citetalias{sysoliatina21} contains a typo; 
parameters $\eta$ and $\zeta$ should be interchanged.} in \citetalias{sysoliatina21}). 
The summation is performed 
over all $N_\mathrm{p}$ peaks. We note that this process of adding extra peaks to the SFR
does not result in an increase of the overall present-day thin-disk surface density 
in the model. Instead, the SFR is re-normalised, such that the total surface density 
still equals to $\Sigma_\mathrm{d}$. 

We can also introduce quantities $f_\mathrm{k}$ and $f_0$ defined as 
contributions of each extra SFR peak and the continuum part to the total  
SFR of a sub-population:
\begin{equation}
f_\mathrm{k}(t) = sfr_\mathrm{dp,k}(t)/sfr_\mathrm{d}(t), 
\ \quad f_0(t) = 1 - \sum_k^{N_\mathrm{p}} f_\mathrm{k}(t). 
\label{eq:fk}
\end{equation}
Here $sfr_\mathrm{dp,k}(t)$ corresponds to the $k$-th term 
of \mbox{Eq. (\ref{eq:sfrdp})}. These fractions are later linked to 
the disk kinematics \mbox{(Section \ref{sect:thindisk_avr})}. 

\begin{figure}[t]
\includegraphics[width=\columnwidth]{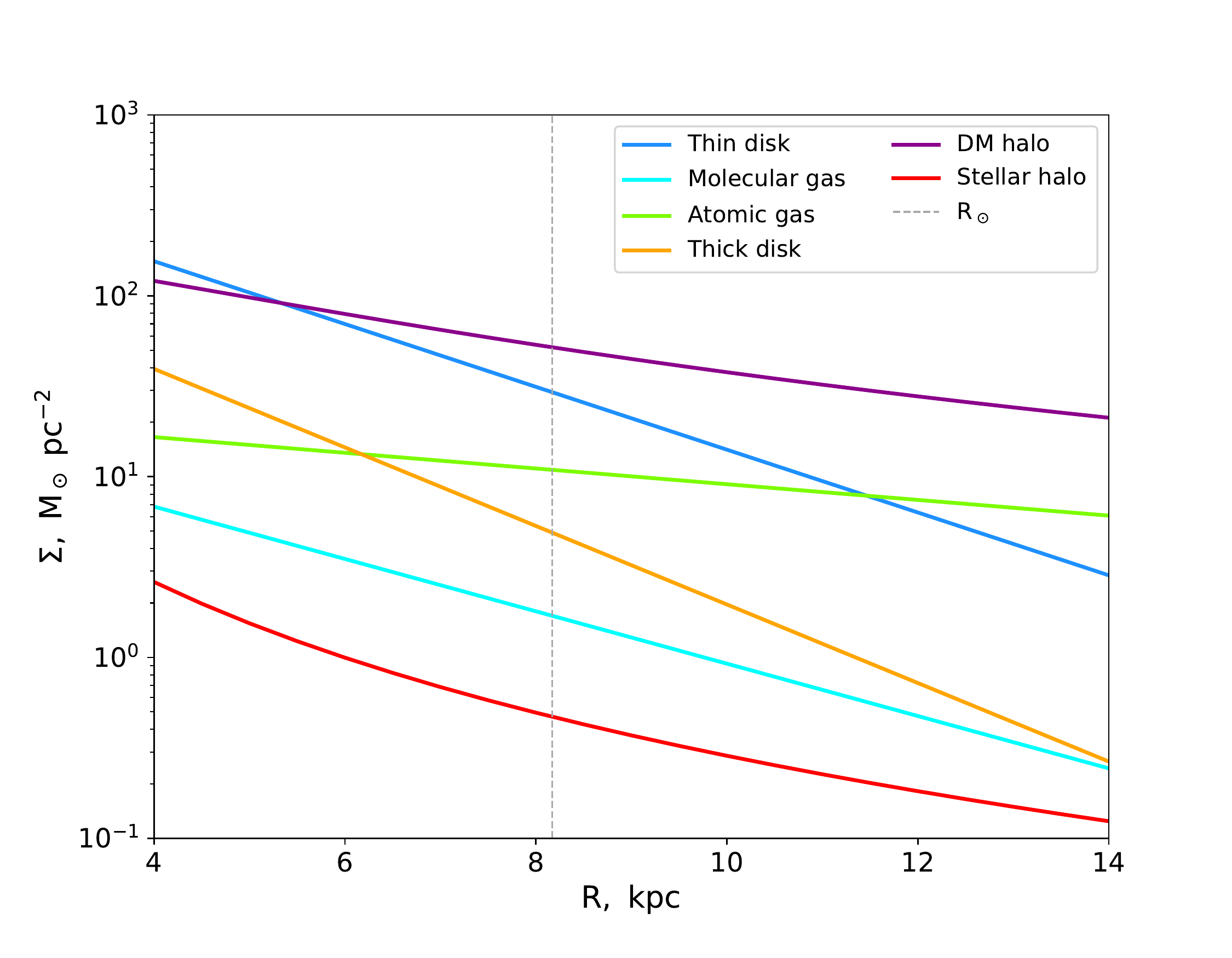}
\caption{Radial density profiles of the JJ-model components. 
The dashed grey line marks the solar position.}
\label{fig:rhor}
\end{figure}

To introduce the radial variation of SFR, we allow some of its parameters 
to be functions of $R$. 
For the radial surface density profile, we assume an exponential law 
with a scale length $R_\mathrm{d}$ laying in the range of \mbox{$\sim$2.5--3 kpc} 
(\mbox{Figure \ref{fig:rhor}}):
\begin{equation}
    \label{eq:disk_sigma}
    \Sigma_\mathrm{d}(R) =  \Sigma_\mathrm{d\odot} 
                \exp{\left( - \frac{R - R_\odot}{R_\mathrm{d}}\right)},
\end{equation}
where $\Sigma_\mathrm{d\odot} = \Sigma_\mathrm{d}(R_\odot)$ is the local surface 
density of the thin disk\footnote{As in the previous papers of this series 
we only considered the local volume, all model parameters were by default local and no 
subscript was used to indicate that. Now we explicitly mark local parameters by
`$\odot$' sign to distinguish between the values in the solar neighbourhood and 
at other galactocentric distances.}. 

Parameter $t_\mathrm{d1}$ is a time delay of SF and can be used to construct the model 
with non-overlapping in time thin- and thick-disk phases. Currently, 
we set this parameter 
to zero at all radii, this corresponds to the in-parallel formation 
of the thin and thick disk. 
Parameters $t_\mathrm{d2}$, $\zeta$, and $\eta$ are now calibrated only 
at the solar radius $R\mathrm{_\odot}$. As a first guess, we 
assumed for them a power-law radial dependence -- a simple assumption 
that at the same time allows the inside-out disk growth to be mimicked. 

We also assume that the thin-disk sub-populations corresponding to each of $N_\mathrm{p}$ peaks form an overdensity normally distributed around some 
mean radii $R_\mathrm{p}$ with dispersions $dR_\mathrm{p}$. Then, 
at an arbitrary radius, the amplitude parameter of the $k$-th SFR peak 
can be calculated as 
\begin{equation}
\Sigma_\mathrm{p,k}(R) = \Sigma_\mathrm{p\odot,k}
\exp{\left( -\frac{R^2 - R_\odot^2 - 2 R_\mathrm{p,k}(R - R_\odot)}
                  {2 \, (dR_\mathrm{p,k})^2} \right)},
\label{eq:Sigmapr}
\end{equation}
where $\Sigma_\mathrm{p\odot}$ is the local value of the parameter. 

\begin{figure}[t]
\includegraphics[width=\columnwidth]{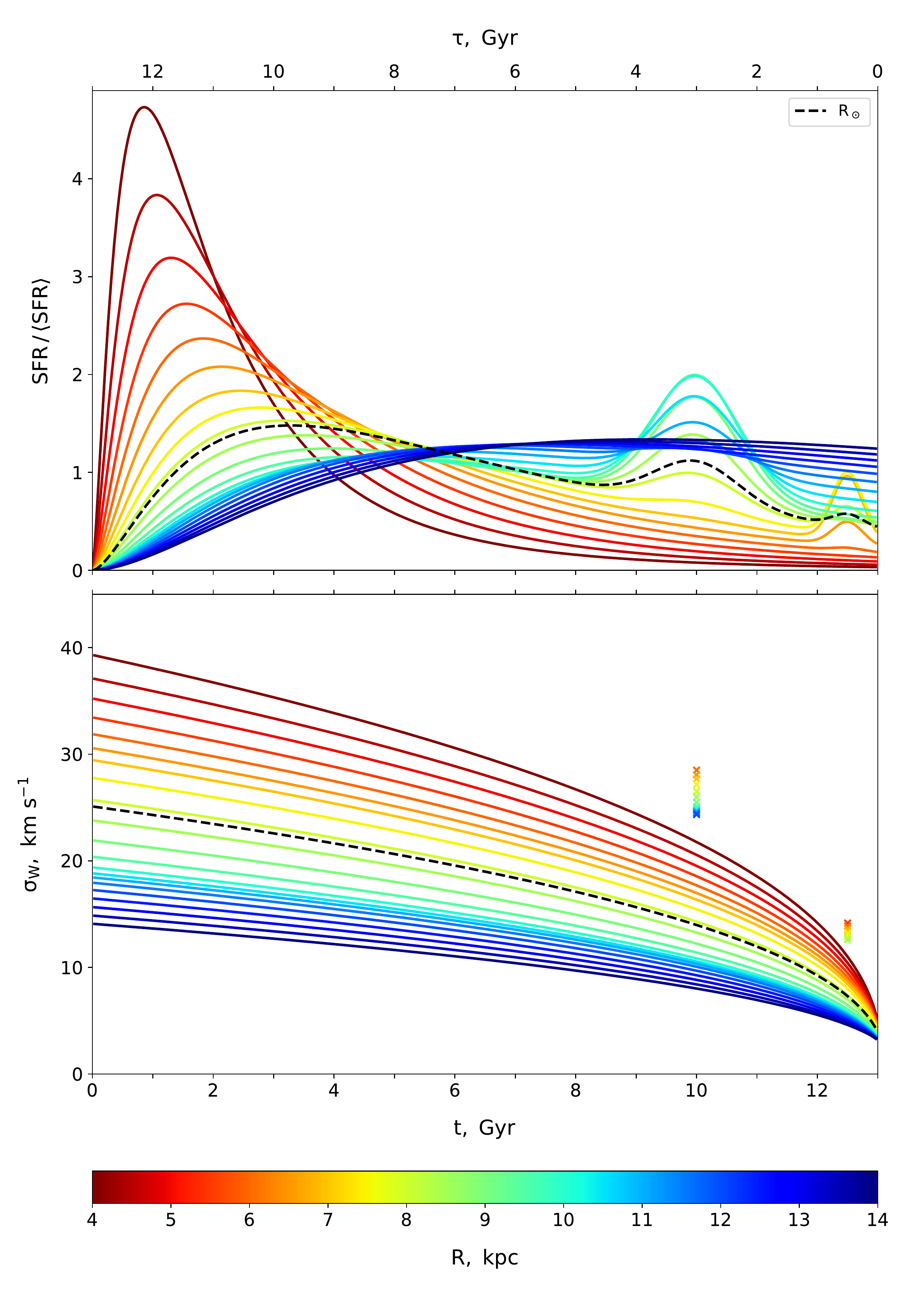}
\caption{Thin-disk SFR (top) and AVR (bottom) as functions of radius. 
Dashed black curves correspond to the solar neighbourhood. 
Crosses give the vertical velocity dispersion corresponding to the fractions 
of sub-populations associated with the extra SFR peaks. 
For the details of this model realisation, see \mbox{Section \ref{sect:model1}}.}
\label{fig:nsfr-avr}
\end{figure}

An example of the radially dependent SFR of the thin disk 
with two additional peaks is shown at the top panel of \mbox{Figure \ref{fig:nsfr-avr}}
(see also \mbox{Section \ref{sect:model1}}). 

\subsubsection{AVR}\label{sect:thindisk_avr}

The vertical kinematics of the thin-disk sub-populations is represented by 
a power-law AVR (\mbox{Eq. (30)} in \citetalias{just10}). At a fixed radius, 
we prescribe it as a product of a normalisation constant, 
scaling parameter $\sigma_\mathrm{e}$, and an age-dependent function, such that  
$\sigma_\mathrm{W}(\tau)  = \sigma_\mathrm{e} \cdot avr(\tau)$. 
We also link the present-day thin-disk velocity dispersion to kinematics of 
the molecular gas (\mbox{Eq. (14)} in \citetalias{sysoliatina21}). 

Following \citetalias{sysoliatina21}, we allow some of the thin-disk sub-populations to have 
kinematics different from the one prescribed by the AVR. More precisely, 
$f_\mathrm{k}$ fractions of sub-populations associated with the $k$-th 
SFR peak (\mbox{Eq. (\ref{eq:fk})}) can have 
the vertical velocity dispersion $\sigma_\mathrm{p,k}$ higher than the one 
prescribed by the AVR function for their age. The remaining  
$f_0$ fractions of the thin-disk sub-populations follow the AVR law. 

To add the radial dependence, we allow one AVR parameter, 
its scaling factor $\sigma_\mathrm{e}$, to vary with $R$. Thus, the AVR shape 
is constant at all radii and is the same as in the solar neighbourhood. 
The value of $\sigma_\mathrm{e}(R)$ 
is iteratively adapted at each radius to obtain the thin disk of a prescribed thickness
(\mbox{Section \ref{sect:scheme}}). 
By default, the thin-disk thickness is constant 
at all radii, which is consistent with observations of edge-on galaxies 
\citep{vanderkruit81,bizyaev02}. 
However, a flaring thin disk can be also tested with 
\mbox{the \texttt{jjmodel} code}, for that a simple exponential flaring model 
from \citet{kalberla14} is included. In this case, the thin-disk half-thickness 
is described by two parameters: radius at which the flaring starts, $R_\mathrm{f}$, and the flaring scale length $R_\mathrm{df}$. 

If the thin-disk SFR is modelled with additional peaks and the peaks-related 
sub-populations are allowed to have special vertical kinematics, then 
at an arbitrary radius, velocity dispersion corresponding to the $k$-th 
SFR peak is calculated as follows:
\begin{equation}
\label{eq:sigmapr}
 \sigma_\mathrm{p,k}(R) = \sqrt{\sigma_\mathrm{ex,k}^2  +      \sigma_\mathrm{W}(R,\tau_\mathrm{p,k})^2},
\end{equation}
\begin{equation*}
 \text{where} \quad 
 \sigma_\mathrm{ex,k}^2 =  \sigma_\mathrm{p\odot,k}^2 -  \sigma_\mathrm{W}(R_\odot,\tau_\mathrm{p,k})^2. 
\end{equation*}
Here $\sigma_\mathrm{p\odot,k}$ is the local value of the parameter, and  
$\sigma_\mathrm{ex,k}$ represents the local velocity dispersion excess 
with respect to the AVR. The value of this excess is then quadratically added 
to the AVR velocity dispersion at each radius. 

The bottom panel of \mbox{Figure \ref{fig:nsfr-avr}} shows AVR corresponding to 
the JJ model with the thin disk of a constant thickness and 
SFR from the upper panel (\mbox{Section \ref{sect:model1}}). 
Crosses mark velocity dispersion of the SFR peaks' populations.

\ifx
\begin{align}
\rho_\mathrm{d} &= f_0 \sum_j^{N_\mathrm{d}} 
  \rho_{0,j} \exp{(-\Phi/\sigma_\mathrm{d,j}^2)} + \\ \nonumber
  &+ \sum_k^{N_\mathrm{p}} f_k \sum_j^{N_\mathrm{d}} \rho_\mathrm{0,j} \exp{(-\Phi/\sigma_\mathrm{p,k}^2)}
\label{eq:rhodp}
\end{align}
\fi

\subsection{Thick disk}\label{sect:thickdisk}

The thick disk is modelled in the full analogy to the thin disk, but 
is described by a smaller number of parameters.

In \mbox{the JJ model}, the thick disk is constructed out of a set of 
mono-age locally isothermal sub-populations covering the first several Gyr 
of the MW evolution. The thick-disk SFR is a power-exponential function 
that peaks at old ages, \mbox{$\sim$12.5 Gyr} ago 
(Eqs. (7) and (8) and Table 2 in \citetalias{sysoliatina21}). 
Following \citetalias{sysoliatina21}, we adopt here 
\mbox{$t_\mathrm{t2}=4$ Gyr} for the duration of the thick-disk formation, 
which corresponds to $N_\mathrm{t} = 160$ mono-age sub-populations. 
At a fixed radius, all thick-disk sub-populations have the same 
vertical velocity dispersion $\sigma_\mathrm{t}$, such that kinematically 
the thick disk behaves as a single isothermal sub-population with the total surface 
density $\Sigma_\mathrm{t}$. 
The thick disk is also characterised by its own AMR function in a form of 
$\tanh$-law (Eqs. (\ref{eq:amr_fit}) and (\ref{eq:f_amrt}) in 
\mbox{Section \ref{sect:amr_fit}}). 

When modelling the thick disk at the different radii, we do not vary the shape 
of its SFR, but only normalise it to the corresponding surface density 
$\Sigma_\mathrm{t}(R)$. For the radial profile of the thick-disk surface 
density we assume, 
by analogy to \mbox{Eq. (\ref{eq:disk_sigma})}, an exponential decline 
with a scale length shorter than that of the thin disk, 
$R_\mathrm{t} \approx 2$ kpc. The thick disk is prescribed to have  
a constant thickness, which we use as a constraint to calculate its velocity 
dispersion $\sigma_\mathrm{t}(R)$ at the radii other than solar. 
The thick-disk AMR is also not varied with galactocentric distance in our 
model, which is consistent with the well-mixed thick-disk population 
observed in our Galaxy (e.g. \citealp{hayden15,hayden20}). 

When the \mbox{JJ model} is used to model volumes extending to heights 
larger than \mbox{$z_\mathrm{max} = 2$ kpc} (e.g. a cone perpendicular to 
the Galactic plane in \citetalias{just11} and \citetalias{sysoliatina21}),
we extrapolate the derived thick-disk vertical profile 
with a \mbox{$\sech^\alpha$-law} of a single isothermal sub-population.
As we showed in \citetalias{just11}, a simple \mbox{$\sech^\alpha$-law} 
is still well-suited for the fitting of the thick-disk vertical profile 
calculated in the presence of DM and other stellar components. 

\subsection{Gas}\label{sect:gas}
Our current gas model for the solar neighbourhood is described in \citetalias{sysoliatina21}, 
which we now extend to other galactocentric distances. 

We consider two gas populations: a cold molecular (H$_2$) and a warm 
atomic (HI) component. The adopted local surface densities of the molecular 
and atomic gas are taken from \citet{mckee15} and are 
\mbox{$\Sigma_\mathrm{H_2}=1.7$ M$_\odot$ pc$^{-2}$} and 
\mbox{$\Sigma_\mathrm{HI}=10.9$ M$_\odot$ pc$^{-2}$}, 
respectively (Table \ref{tab:jjparams} and 
\mbox{Table 2} in \citetalias{sysoliatina21}). 
For the surface density radial profiles of both gas components, 
we assume exponential laws. \citet{kramer16} discuss the radial column and 
surface density profiles of the atomic and molecular gas as derived by 
different authors before \mbox{2006--2008 yrs}. 
An averaged surface density profile of the molecular gas peaks at \mbox{$\sim$4 kpc} 
and then declines with a short scale length of \mbox{$\sim$2--3 kpc}. 
We now adopt \mbox{$R_\mathrm{H_2}=3$ kpc} as a default value of the molecular 
gas scale length in the generalised \mbox{JJ model}. 
In the case of atomic gas, its surface density remains roughly constant in 
the wide distance range of \mbox{4--14 kpc} and declines with a scale length of 
\mbox{$\sim$4.5 kpc} in the outer disk. Similarly, \citet{kalberla08} describe 
HI radial distribution: it is approximately constant in the inner disk and 
exponentially declining in the outer disk with a scale length of 
\mbox{$\sim$3.75 kpc} at \mbox{12.5 kpc} $< \mathrm{R} <$ \mbox{30 kpc}.
We input \mbox{$R_\mathrm{HI}=10$ kpc} to describe this slow decline 
of the atomic gas surface density with radius, 
which is broadly consistent with the mentioned observational data. 

By analogy to the thin-disk mono-age populations, both gas components 
are treated as isothermal at each radius. Their vertical velocity dispersions 
are adapted to reproduce the prescribed scale heights of the cold and warm gas. 
We adopt the exponentially growing radial profiles of the H$_2$ 
and HI half-thickness from \citet{nakanishi16}, which 
the authors measured using their three-dimensional gas maps constructed on the basis of  
HI and CO surveys data and kinematic distances to gas clouds. 

In order to get a realistic enclosed mass of gas for  
the MW rotation curve, we add inner holes for both gas layers 
at $R_\mathrm{0,H_2}$ and $R_\mathrm{0,HI}$, 
which roughly reflects a sharp decrease of gas density in the inner disk. 

\subsection{Stellar halo}\label{sect:stellarhalo}

Stellar halo is modelled as a globally isothermal population 
of a single age $t_\mathrm{p}$. 
At each radius, its vertical density profile depends on two parameters, 
a universal vertical velocity dispersion $\sigma_\mathrm{sh}$ and 
its surface density at this radius $\Sigma_\mathrm{sh}(R)$. 
To get the radial surface density profile of the halo, we numerically integrate 
its power-law matter density profile up to $z_\mathrm{max}$. The 
default power-law slope is $\alpha_\mathrm{in} = -2.5$ corresponding 
to the inner halo \citep{bhawthorn16}:
\begin{align}
\label{eq:sh_sigma}
\Sigma_\mathrm{sh}(R) &= \frac{\Sigma_\mathrm{sh \odot}}{C}
        \int_0^{z_\mathrm{max}}
\left(\frac{r(R,z)}{r(R_\odot,z_\odot)}\right)^{-\alpha_\mathrm{in}} dz, 
\quad \text{where} \\
C &= \int_0^{z_\mathrm{max}}
\left(\frac{r(R_\odot,z)}{r(R_\odot,z_\odot)}\right)^{-\alpha_\mathrm{in}} dz. \nonumber
\end{align}
Here $\Sigma_\mathrm{sh\odot}$ is the local surface density of the halo, 
and $C$ is a normalisation constant, and $r(R,z) = \sqrt{R^2 + z^2}$ 
is a radial galactocentric coordinate. 

Here we do not include halo flattening, as it will introduce only a small correction to 
\mbox{Eq. (\ref{eq:sh_sigma})}. As the halo itself makes the smallest contribution 
to the total density among all other components, 
adding its flattening can be viewed as a second-order correction
to the model predictions, so we neglect it for now. Though, in some cases 
more comprehensive modelling of the halo is justified. In \citetalias{sysoliatina21}, 
we simulated a vertically extended sample consisting of \textit{Gaia} DR2 stars 
selected in a cone 
perpendicular to the Galactic plane. To get the right halo star counts, we extended the 
vertical halo profile that was derived self-consistently with potential only up to
$z_\mathrm{max}$.
For extending it to larger $z$, we used a broken power-law profile representing the inner 
and outer halo (Eq. (15) in \citetalias{sysoliatina21}). If necessary, flattening can be also added at this stage.

When creating mock samples, we assume for our single-age halo 
a Gaussian MD with a mean $\mathrm{[Fe/H]}_\mathrm{sh}$
and a dispersion $d\mathrm{[Fe/H]}_\mathrm{sh}$. That is to say, 
the halo population is further 
subdivided into $n_\mathrm{sh} = \,$5--9 sub-populations with different 
metallicities (see also \mbox{Section \ref{sect:sps}}).

\subsection{DM halo}\label{sect:dmhalo}

The DM halo is treated in the \mbox{JJ model} similarly to the stellar halo. 
At each radius, DM component is added in the form a single isothermal population 
characterised by the vertical velocity dispersion $\sigma_\mathrm{dh}$, which is 
constant at all radii, and a radially dependent surface density 
$\Sigma_\mathrm{dh}(R)$.

Assuming for the spatial DM density a cored isothermal sphere, 
we can express the radial DM surface density as follows: 
\begin{equation}
    \label{eq:dm_sigma}
    \Sigma_\mathrm{dh}(R) = \Sigma_\mathrm{dh\odot} 
                    \frac{r_\mathrm{a_h}(R_\odot)}{r_\mathrm{a_h}(R)} 
            \frac{\arctan{\left(z_\mathrm{max}/r_\mathrm{a_h}(R)\right)}}
                 {\arctan{\left(z_\mathrm{max}/r_\mathrm{a_h}(R_\odot) \right)}}, 
\end{equation}
where $\Sigma_\mathrm{dh\odot}$ is the local DM surface density,  
$a_\mathrm{h}$ is the DM scaling parameter, and 
$r_\mathrm{a_h}(R) = \sqrt{a_\mathrm{h}^2 + R^2}$. 
The scaling parameter is adapted to have 
the local total circular velocity $\upsilon_\mathrm{c,\odot}$ consistent with 
the observed proper motion of \mbox{Sgr A$^\ast$} \citep{reid05} 
at the assumed solar radius $R_\odot$, 
with the peculiar motion of the Sun $V_\odot$ taken into account. 

\subsection{Bulge}\label{sect:bulge}

Though we do not model the inner \mbox{4 kpc} of the MW disk, 
the bulge presence needs to be taken into account for the calculation 
of the rotation curve. As we mentioned above in \mbox{Section \ref{sect:dmhalo}}, 
we use the predicted circular velocity at the solar radius to constrain 
the DM scaling parameter $a_\mathrm{h}$, which then enters 
\mbox{Eq. (\ref{eq:dm_sigma})}. For this purpose, we include the bulge 
in a simple form of a point mass characterised by a single parameter, 
mass $M_\mathrm{b}$. As we do not add an inner hole in the thin- and thick-disk 
radial density profiles, $M_\mathrm{b}$ gives a difference between 
the bulge mass and the enclosed mass of our exponential disks extrapolated 
to the inner MW region.

\subsection{Creating mock samples}\label{sect:sps}

In order to convert modelled density profiles into the mock samples of  
some observable populations, we need to use a stellar evolution library. 
As we demonstrated in \citet{sysoliatina18} and \citetalias{sysoliatina21}, 
the choice of the stellar library may have a significant impact on the 
location of different features on the colour-magnitude diagram (CMD) 
and on the overall predicted star counts introducing 
up to \mbox{$\sim$10-\%} uncertainty. To estimate 
the impact of the stellar library choice on our modelling, we used 
in our previous works different 
sets of isochrones that we now include in the \texttt{jjmodel} 
% A&A version
\ifx
package. 

First, our default set of isochrones is generated with the  
PAdova and TRieste Stellar Evolution Code (PARSEC v.1.2S; \citealp{bressan12})
and the COLIBRI code (v. S{\_}35; \citealp{marigo17}) 
for the thermally pulsing asymptotic giant branch  
phase\footnote{\url{http://stev.oapd.inaf.it/cgi-bin/cmd}}. 
Our isochrone grid includes photometric columns in the $UBVRIJHK$ system 
and $G_\mathrm{BP}$, $G_\mathrm{RP}$, and $G$ bands 
for \mbox{the \textit{Gaia} DR2} \citep{maiz-apellaniz18} and 
EDR3 \citep{riello21}. 

Second, a complementary isochrone grid is based on \mbox{the MESA}
(Modules and Experiments in Stellar Astrophysics,
\mbox{\citealp{paxton11,paxton13,paxton15}}) Isochrones and Stellar Tracks 
(MIST\footnote{\url{http://waps.cfa.harvard.edu/MIST/}}, v.1.2;
\citealp{dotter16,choi16}). For this stellar library, 
the following photometric bands are provided in the \texttt{jjmodel} code:  
standard $UBVRI$, 2 Micron All Sky Survey (2MASS; \citealp{skrutskie06}) 
$JHK_s$, and three \textit{Gaia} $G$-bands 
for DR2 and EDR3. 

Third, another complementary grid is based on the Bag of Stellar Tracks and 
Isochrones (BaSTI\footnote{\url{http://basti-iac.oa-abruzzo.inaf.it/isocs.html}},
\citealp{hidalgo18}). The available photometric system is \textit{Gaia} EDR3.
\fi
% Version for arxiv with a list
package. 
\begin{itemize}
    \item Our default set of isochrones is generated with the  
    PAdova and TRieste Stellar Evolution Code (PARSEC v.1.2S; \citealp{bressan12})
    and the COLIBRI code (v. S{\_}35; \citealp{marigo17}) 
    for the thermally pulsing asymptotic giant branch  
    phase\footnote{\url{http://stev.oapd.inaf.it/cgi-bin/cmd}}. 
    Our isochrone grid includes photometric columns in the $UBVRIJHK$ system 
    and $G_\mathrm{BP}$, $G_\mathrm{RP}$, and $G$ bands 
    for \mbox{the \textit{Gaia} DR2} \citep{maiz-apellaniz18} and 
    EDR3 \citep{riello21}. 
    
    \item A complementary isochrone grid is based on \mbox{the MESA}
    (Modules and Experiments in Stellar Astrophysics,
    \mbox{\citealp{paxton11,paxton13,paxton15}}) Isochrones and Stellar Tracks 
    (MIST\footnote{\url{http://waps.cfa.harvard.edu/MIST/}}, v.1.2;
    \citealp{dotter16,choi16}). For this stellar library, 
    the following photometric bands are provided in the \texttt{jjmodel} code:  
    standard $UBVRI$, 2 Micron All Sky Survey (2MASS; \citealp{skrutskie06}) 
    $JHK_s$, and three \textit{Gaia} $G$-bands 
    for DR2 and EDR3. 
    
    \item Another complementary grid is based on the Bag of Stellar Tracks and 
    Isochrones (BaSTI\footnote{\url{http://basti-iac.oa-abruzzo.inaf.it/isocs.html}};
    \citealp{hidalgo18}). The available photometric system is \textit{Gaia} EDR3.
\end{itemize}

Following \citetalias{sysoliatina21}, we use a wide-range metallicity grid
\mbox{$-2.6...+0.47$} of 62 isochrone tables, each covering 
the whole modelled age interval \mbox{$0...t_\mathrm{p}$} with a linear step 
of \mbox{$50$ Myr}. Each of the \mbox{JJ-model} mono-age isothermal
sub-populations of the thin disk, thick disk, and stellar halo is characterised 
by an age-metallicity pair -- according to the disks' AMR 
\mbox{(Sections \ref{sect:thindisk_amr} and \ref{sect:thickdisk})} 
and the halo MD \mbox{(Section \ref{sect:stellarhalo})}.
In the case of the thin and thick disk sub-populations, 
a spread in metallicity can be additionally allowed: for each age, we assumed 
a Gaussian MD centred at the metallicity prescribed for 
this age by the AMR function, $\mathrm{[Fe/H]}(\tau_\mathrm{j})$,
and characterised by dispersion $d\mathrm{[Fe/H]}_\mathrm{dt}$, 
which is the same for all ages. 
In practice, this means that we increase the number of modelled age-metallicity pairs 
by the number of sub-populations used to sample the assumed MD 
at each age, $n_\mathrm{dt}$. Though, this added complexity slows down 
the calculation process, it can help to mimic the effect of radial migration 
and produce more realistic Hess diagrams \citep{sysoliatina18}. 
In the end, the total number of the stellar mono-age mono-metallicity sub-populations
can be calculated as  
$(N_\mathrm{d} + N_\mathrm{t})\cdot n_\mathrm{dt} + n_\mathrm{sh}$, 
where $n_\mathrm{dt} = 1$ if spread in metallicity for the disks is not allowed. 

For each mono-age mono-metallicity sub-population, we select the best isochrone 
(the one closest in metallicity and age). 
This isochrone then provide us with a set of `stellar assemblies' (SA) -- 
populations 
of the same age, metallicity, and mass. The exact SA number, $N_\mathrm{SA}$,
corresponding to a single isochrone depends on metallicity, age, and most of all,
on the mass resolution of the stellar library. 
On average, $N_\mathrm{SA}$ for a fixed radius is of the order of $10^5$--10$^6$
in \mbox{the JJ model}. 

At the next step, we calculate for each SA its present-day surface number 
density according to the adopted IMF:
\begin{equation}
N^\Sigma_\mathrm{l} = \frac{dn}{dm}(m^{low}_\mathrm{l},m^{up}_\mathrm{l},
                        M_\mathrm{l}).
\label{eq:Nsigma}
\end{equation}
Here $m^{low}_\mathrm{l}$ and $m^{up}_\mathrm{l}$ are the boundaries of 
the SA mass range, and $M_\mathrm{l}$ 
is the overall mass converted into stars at the time interval corresponding 
to the SA age $\tau_\mathrm{l}$. 
For the thin and thick disk, this mass is given by their SFR 
multiplied by the time resolution, 
$M_\mathrm{l}= SFR(\tau_\mathrm{l}) \, t_\mathrm{r} \, \omega_\mathrm{l}$.  
Here $\omega_\mathrm{l}=1$ if there is no metallicity spread allowed for the 
mono-age sub-populations or equals to a corresponding Gaussian weight otherwise. 
For a single-age halo for which we do not have SFR, the initial mass 
converted into stars is calculated by dividing its present-day surface density 
by its mass loss factor, 
$M_\mathrm{l} = \Sigma_\mathrm{sh}/g_\mathrm{sh} \cdot \omega_{l}$. 
The adopted fraction of the halo mass lost due to stellar evolution, 
$g_\mathrm{sh}$, corresponds to the adopted mean halo metallicity
$\mathrm{[Fe/H]}_\mathrm{sh}$ and is driven 
from our pre-calculated mass-loss grid based on the \textit{Chempy} code 
(\mbox{Section \ref{sect:thindisk}}). By analogy to the thin and thick disk, 
weights $\omega_{l}$ are Gaussian weights that originate from the halo MD sampling. 

Following this scheme at each radius, we create SA tables for the thin disk, 
thick disk, and halo and store them for later use. As isochrones provide us with 
a set of useful stellar parameters (such as present-day stellar mass $M_\mathrm{f}$, luminosity $\log{L}$, effective temperature $\log{T_{\mathrm{eff}}}$, and 
surface gravity $\log{g}$) and absolute magnitudes in different photometric bands, 
we can use this information to define observable samples 
(e.g. see the selection of the RC stars in \mbox{Section \ref{sect:RC_in_model}}). 
For the SA subset corresponding to a mock population, we then calculate spatial number densities as a function of height at a fixed $R$: 
\begin{equation}
N^V_\mathrm{l}(|z|) = \frac{N^{\Sigma}_\mathrm{l}}{2 h(\tau_\mathrm{l})}
\exp{\left(- \Phi(|z|)/\sigma_\mathrm{W}^2(\tau_\mathrm{l})\right)}.
\label{eq:Nzv}
\end{equation}
Here $h(\tau_\mathrm{l})$ and $\sigma_\mathrm{W}^2(\tau_\mathrm{l})$ are 
the SA scale height determined during an iterative solving of the Poisson equation \mbox{(Section \ref{sect:scheme})} and its vertical velocity dispersion.
For the thin disk, $\sigma_\mathrm{W}$ depends on age and is given by the AVR 
function. 
We note that if extra $N_\mathrm{p}$ peaks are added to the thin-disk SFR, 
and the corresponding sub-populations are allowed to have special vertical kinematics 
(\mbox{Sections \ref{sect:thindisk_sfr} and \ref{sect:thindisk_avr}}), 
the right-hand side of \mbox{Eq. (\ref{eq:Nzv})}  
splits into $N_\mathrm{p} + 1$ terms. In the first term, which is weighted 
by the factor $f_0$ (\mbox{Eq. (\ref{eq:fk})}), $\sigma_\mathrm{W}$ is 
prescribed by the AVR. In the rest $N_\mathrm{p}$ terms, 
weighted by factors $f_\mathrm{k}$, $\sigma_\mathrm{W}$ is the velocity 
dispersion of the $k$-th peak, $\sigma_\mathrm{p,k}$. 
The application of \mbox{Eq. (\ref{eq:Nzv})} to the thick-disk and halo SA 
is much simpler. In this case, the role of $\sigma_\mathrm{W}$ play  
quantities $\sigma_\mathrm{t}$ and $\sigma_\mathrm{sh}$. Also, velocity dispersions
and scale heights of all thick-disk and halo SAs are independent of age
by construction of our model. 

With \mbox{Eq. (\ref{eq:Nzv})}, prediction of star counts in some volume 
is straightforward. 
If the volume radial extension exceeds the adopted radial resolution of 
\mbox{the JJ model}, it can be split into several radial slices, in each of which 
\mbox{Eq. (\ref{eq:Nzv})} has to be applied separately.

\subsection{Modelling scheme}\label{sect:scheme}

\begin{table*}[ht]
    \centering
    \caption{Parameters of the test generalised \mbox{JJ model} $\mathrm{TM1}$. 
    The thin- and thick-disk AMR parameters are listed separately in 
    \mbox{Table \ref{tab:amr_params}}. 
    Parameters in this table can be divided into the following categories: 
    (a) calibrated against Gaia DR2 in \citetalias{sysoliatina21}, 
    (b) calibrated against APOGEE RC in \citetalias{sysoliatina21}, 
    (c) adopted in \citetalias{sysoliatina21}, 
    (d) adopted in \citetalias{just10},
    (e) consistent with values from the literature (see text), 
    (f) not calibrated.}
    \begin{tabular}{l|l} \hhline{==}
    
    {\begin{minipage}[c][0.6cm][c]{2.5cm} solar parameters \end{minipage}} &
        $R_\odot = 8.17$ kpc$\, ^{(e)}$, 
        $z_\odot = 20$ pc$\, ^{(e)}$, 
        $V_\odot = 12.5$  km s$^{-1} \, ^{(e)}$ \\ \hline
        
    {\begin{minipage}[c][0.6cm][c]{2.5cm} Spatial grid \end{minipage}} & 
        $R_\mathrm{min} = 4$ kpc, 
        $R_\mathrm{max} = 14$ kpc,
        $dR = 0.5$ kpc, 
        $z_\mathrm{max} = 2000$	pc, 
        $dz = 2$ pc \\ \hline
        
    {\begin{minipage}[c][1.2cm][c]{2.5cm} Radial profiles \end{minipage}} & 
        {\begin{minipage}[c][1.2cm][c]{12.5cm}
            $R_\mathrm{d} = 2.5$ kpc$\, ^{(e)}$, 
            $R_\mathrm{t} = 2$ kpc$\, ^{(e)}$,
            $R_\mathrm{H_2} = 3$ kpc$\, ^{(e)}$, 
            $R_\mathrm{HI} = 10$ kpc$\, ^{(e)}$, \\
            $R_\mathrm{0,H_2} = 4$ kpc$\, ^{(e)}$, 
            $R_\mathrm{0,HI} = 4$ kpc$\, ^{(e)}$, 
            $\alpha_\mathrm{in} = -2.5\, ^{(e)}$, 
            $M_\mathrm{b} = 0.8 \cdot 10^{10}$ M$_\odot \, ^{(e)}$ 
        \end{minipage}} \\ \hline
    
    {\begin{minipage}[c][1.2cm][c]{2.5cm} Local \\ normalisations \end{minipage}} &
        {\begin{minipage}[c][1.2cm][c]{12.5cm} 
            $\Sigma_\mathrm{d\odot} = 29.4$ M$_\odot$ pc$^{-2} \, ^{(a)}$, 
            $\Sigma_\mathrm{t\odot} = 4.9$ M$_\odot$ pc$^{-2} \, ^{(a)}$, 
            $\Sigma_\mathrm{H_2\odot} = 1.7$ M$_\odot$ pc$^{-2} \, ^{(e)}$, \\
            $\Sigma_\mathrm{HI\odot} = 10.9$ M$_\odot$ pc$^{-2} \, ^{(e)}$, 
            $\Sigma_\mathrm{dh\odot} = 51.9$ M$_\odot$ pc$^{-2} \, ^{(a)}$,
            $\Sigma_\mathrm{sh\odot} = 0.47$ M$_\odot$ pc$^{-2} \, ^{(a)}$ 
        \end{minipage}} \\ \hline
            
    {\begin{minipage}[c][2.4cm][c]{2.5cm} SFR \end{minipage}} & 
        {\begin{minipage}[c][2.4cm][c]{12.5cm} 
            $t_\mathrm{d1\odot} = 0 $ Gyr$\, ^{(c)}$, 
            $t_\mathrm{d2\odot} = 7.8 $ Gyr$\, ^{(c)}$, 
            $\zeta_\odot = 0.83\, ^{(a)}$, 
            $\eta_\odot = 5.6\, ^{(a)}$, \\
            $k_\mathrm{t_{d2}} = 1.5 \, ^{(f)}$, 
            $k_\mathrm{\zeta} = 0.18 \, ^{(f)}$, 
            $k_\mathrm{\eta} = -0.1 \, ^{(f)}$, \\
            $\Sigma_\mathrm{p\odot} = [3.5,1.3]$ M$_\odot$ pc$^{-2} \, ^{(a)}$,
            $\tau_\mathrm{p} = [3,0.5]$ Gyr$\, ^{(a)}$, 
            $d\tau_\mathrm{p} = [0.7,0.25]$ Gyr$\, ^{(a)}$, \\
            $R_\mathrm{p} = [9,7]$ kpc$\, ^{(f)}$, 
            $dR_\mathrm{p} = [1,0.5]$ kpc$\, ^{(f)}$ \\
            $t_\mathrm{t1} = 0.1 $ Gyr$\, ^{(c)}$, 
            $t_\mathrm{t2} = 4$ Gyr$\, ^{(c)}$, 
            $\gamma = 2 \, ^{(c)}$, 
            $\beta = 3.5$ Gyr$^{-1} \, ^{(c)}$ 
        \end{minipage}} \\ \hline
            
    {\begin{minipage}[c][1.2cm][c]{2.5cm} IMF \end{minipage}} & 
        {\begin{minipage}[c][1.2cm][c]{12.5cm} 
            $\alpha_0 = 1.31 \, ^{(a)}$, 
            $\alpha_1 = 1.5 \, ^{(a)}$, 
            $\alpha_2 = 2.88 \, ^{(a)}$, 
            $\alpha_3 = 2.28 \, ^{(a)}$, \\
            $m_0 = 0.49$ M$_\odot \, ^{(a)}$, 
            $m_1 = 1.43$ M$_\odot \, ^{(a)}$,
            $m_2 = 6$ M$_\odot \, ^{(e)}$ 
        \end{minipage}} \\ \hline
        
    {\begin{minipage}[c][0.6cm][c]{2.5cm} Metallicities \end{minipage}} &
        {\begin{minipage}[c][0.6cm][c]{12.5cm} 
            $\mathrm{[Fe/H]}_\mathrm{sh} = -1.5\, ^{(e)}$, 
            $d\mathrm{[Fe/H]}_\mathrm{sh} = 0.4\, ^{(e)}$ 
        \end{minipage}} \\ \hline
            
    {\begin{minipage}[c][1.2cm][c]{2.5cm} W-velocities \end{minipage}} & 
        {\begin{minipage}[c][1.2cm][c]{12.5cm} 
            $\alpha = 0.409 \, ^{(a)}$, 
            $\sigma_\mathrm{e\odot} = 25.1$ km s$^{-1} \, ^{(a)}$, 
            $\sigma_{p\odot} = [26.3,12.6]$ km s$^{-1} \, ^{(a)}$, \\
            $\sigma_\mathrm{t\odot} = 43.3$ km s$^{-1} \, ^{(a)}$,  
            $\sigma_\mathrm{dh} = 140$ km s$^{-1} \, ^{(d)}$, 
            $\sigma_\mathrm{sh} = 100$ km s$^{-1} \, ^{(e)}$ 
        \end{minipage}} \\ \hhline{==}
        
    \end{tabular}
    \label{tab:jjparams}
\end{table*}

At this point, we introduced all elements of the \mbox{JJ model}  
(\mbox{Sections \ref{sect:assumptions} -- \ref{sect:sps}}), 
so we can finally summarise the key steps of modelling process. 
To do so, we firstly rehearse physical parameters of \mbox{the JJ model} 
(\mbox{Section \ref{sect:model_params}}), 
and then list the sequence of steps required to create mock samples 
(\mbox{Section \ref{sect:build_disk}}). 

\subsubsection{JJ-model parameters}\label{sect:model_params}

The parameter file of the \texttt{jjmodel} package includes not only 
physical parameters of the model, but also some technical ones. 
The latter are boolean quantities that 
code special instructions (e.g. whether to add extra peaks to the thin-disk SFR 
or not, use a constant-thickness or flaring thin disk, build the local or radially 
extended model, etc.). The complete list of \mbox{the JJ-model} parameters with 
comments on their meaning, units, and usage can be found in 
\mbox{the \texttt{jjmodel}} documentation\footnote{\url{https://github.com/askenja/jjmodel/blob/main/jjmodel/docs/mydoc/JJmodel_documentation.pdf}}. 
Here we discuss only physical parameters, which we divide 
into several groups for convenience (see \mbox{Table \ref{tab:jjparams}}). 
% A&A version
\vspace{3pt}

\textit{Solar parameters.} We specify solar peculiar velocity 
$V_\odot$, its galactocentric distance $R_\odot$, and distance from  
the Galactic plane $z_\odot$. 
\vspace{3pt}

\textit{$R$-$z$ grid.} The spatial grid of \mbox{the JJ model} is given by 
the centres of the inner and outer radial bins, $R_\mathrm{min}$ and 
$R_\mathrm{max}$, bin width $dR$, maximum absolute height 
$z_\mathrm{max}$, and the vertical \mbox{step $dz$}. 
\vspace{3pt}

\textit{Radial profiles.} Another group of parameters prescribe 
the radial structure of the MW components. Scale lengths 
$R_\mathrm{d}$, $R_\mathrm{t}$, $R_\mathrm{H_2}$, and $R_\mathrm{HI}$ describe 
an exponential radial density fall-off of the thin disk, thick disk, 
molecular and atomic gas, respectively. For the gas components, we reserve two 
additional parameters $R_\mathrm{0,H_2}$ and $R_\mathrm{0,HI}$ corresponding 
to the radii of the inner holes in their density profiles. 
Stellar halo radial surface density depends on the inner-halo power-law 
slope $\alpha_\mathrm{in}$, and the point-mass bulge is described by 
its total \mbox{mass $M_\mathrm{b}$}. 
\vspace{3pt}

\textit{Local normalisations.} The overall radial density profiles of 
the modelled MW components are normalised to the local surface densities: 
$\Sigma_\mathrm{d\odot}$, $\Sigma_\mathrm{t\odot}$, 
$\Sigma_\mathrm{H_2\odot}$, $\Sigma_\mathrm{HI\odot}$,
$\Sigma_\mathrm{dh\odot}$, and $\Sigma_\mathrm{sh\odot}$. 
\vspace{3pt}

\textit{SFR.} The universal shape of thick-disk SFR is described by 
four parameters -- $t_\mathrm{t1}$, $t_\mathrm{t2}$, $\gamma$, and $\beta$. 
The continuum part of the local thin-disk SFR is given by $t_\mathrm{d1}$,
$t_\mathrm{d2}$, $\zeta$, and $\eta$. The corresponding power-law indices
$k_\mathrm{t_{d2}}$, $k_\mathrm{\zeta}$, and $k_\mathrm{\eta}$ describe an
extension of the local thin-disk SFR to other galactocentric distances. 
If there are $N_\mathrm{p}$ additional Gaussian peaks added to the standard 
SFR of the thin disk, then each peak is characterised by the following 
five parameters: amplitude-related parameter $\Sigma_\mathrm{p\odot}$, mean age 
$\tau_\mathrm{p}$, age dispersion $d\tau_\mathrm{p}$, mean radius $R_\mathrm{p}$,
and radial dispersion $dR_\mathrm{p}$. 
\vspace{3pt}

\textit{IMF.} Our four-slope broken power-law IMF is defined by seven 
parameters -- four slopes, $\alpha_0$, $\alpha_1$, $\alpha_2$, and $\alpha_3$, 
and three break-point masses, $m_0$, $m_1$, and $m_2$.  
In order to work with a different IMF, one can introduce another 
set of IMF parameters in the \texttt{jjmodel} code.
\vspace{3pt}

\textit{Metallicities.} There are four parameters describing a universal 
thick-disk AMR -- initial and present-day metallicity,
$\mathrm{[Fe/H]}_\mathrm{t0}$ and $\mathrm{[Fe/H]}_\mathrm{tp}$, and parameters 
$t_0$ and $r_\mathrm{t}$ controlling the AMR shape. In the case of 
the thin disk, there are two alternative parameter sets.
If the thin-disk AMR is given by \mbox{Eqs. (21) and (22)} 
in \citetalias{sysoliatina21}, there are four parameters of the local AMR: 
$\mathrm{[Fe/H]}_\mathrm{d0}$, $\mathrm{[Fe/H]}_\mathrm{dp}$, $r_\mathrm{d}$, 
and $q$. Power-law slopes $k_\mathrm{[Fe/H]_{d0}}$, $k_\mathrm{[Fe/H]_{dp}}$,
$k_\mathrm{r_d}$, and $k_\mathrm{q}$ then prescribe the radial change 
of the thin-disk AMR. When the thin-disk AMR is given by 
\mbox{Eqs. (\ref{eq:amr_fit}) and (\ref{eq:f_amrd})}, an alternative set of parameters must be given (first row in \mbox{Table \ref{tab:amr_params}}; 
see \mbox{Section \ref{sect:amr_fit}} for details). 
Parameter $d\mathrm{[Fe/H]_{dt}}$ specifies 
the metallicity dispersion applied to the disks' mono-age sub-populations. 
The halo MD is described by the mean metallicity 
$\mathrm{[Fe/H]}_\mathrm{sh}$ and dispersion $d\mathrm{[Fe/H]}_\mathrm{sh}$. 
\vspace{3pt}

\textit{$W$-velocities.} The thin-disk local kinematics is given by 
the AVR power index $\alpha$ and scaling factor $\sigma_\mathrm{e}$. 
Special kinematics of the sub-populations associated with each of the additional SFR peaks is described by $W$-velocity dispersion $\sigma_\mathrm{p}$. 
The local dispersions of the thick disk, DM, and stellar halo are given by parameters $\sigma_\mathrm{t}$, $\sigma_\mathrm{dh}$, and $\sigma_\mathrm{sh}$, respectively. 
\vspace{3pt}

% Arxiv version
\ifx
\begin{itemize}
\setlength\itemsep{0.25em} 
    \item \textit{Solar parameters.} We specify solar peculiar velocity 
    $V_\odot$, its galactocentric distance $R_\odot$, and distance from  
    the Galactic plane $z_\odot$. 
    
    \item \textit{$R$-$z$ grid.} The spatial grid of \mbox{the JJ model} is given by 
    the centres of the inner and outer radial bins, $R_\mathrm{min}$ and 
    $R_\mathrm{max}$, bin width $dR$, maximum absolute height 
    $z_\mathrm{max}$, and the vertical \mbox{step $dz$}. 
    
    \item \textit{Radial profiles.} Another group of parameters prescribe 
    the radial structure of the MW components. Scale lengths 
    $R_\mathrm{d}$, $R_\mathrm{t}$, $R_\mathrm{H_2}$, and $R_\mathrm{HI}$ describe 
    an exponential radial density falloff of the thin disk, thick disk, 
    molecular and atomic gas, respectively. For the gas components, we reserve two 
    additional parameters $R_\mathrm{0,H_2}$ and $R_\mathrm{0,HI}$ corresponding 
    to the radii of the inner holes in their density profiles. 
    Stellar halo radial surface density depends on the inner-halo power-law 
    slope $\alpha_\mathrm{in}$, and the point-mass bulge is described by 
    its total \mbox{mass $M_\mathrm{b}$}. 
    
    \item \textit{Local normalisations.} The overall radial density profiles of 
    the modelled MW components are normalised to the local surface densities: 
    $\Sigma_\mathrm{d\odot}$, $\Sigma_\mathrm{t\odot}$, 
    $\Sigma_\mathrm{H_2\odot}$, $\Sigma_\mathrm{HI\odot}$,
    $\Sigma_\mathrm{dh\odot}$, and $\Sigma_\mathrm{sh\odot}$. 
    
    \item \textit{SFR.} The universal shape of thick-disk SFR is described by 
    four parameters -- $t_\mathrm{t1}$, $t_\mathrm{t2}$, $\gamma$, and $\beta$. 
    The continuum part of the local thin-disk SFR is given by $t_\mathrm{d1}$,
    $t_\mathrm{d2}$, $\zeta$, and $\eta$. The corresponding power-law indices
    $k_\mathrm{t_{d2}}$, $k_\mathrm{\zeta}$, and $k_\mathrm{\eta}$ describe an
    extension of the local thin-disk SFR to other galactocentric distances. 
    If there are $N_\mathrm{p}$ additional Gaussian peaks added to the standard 
    SFR of the thin disk, then each peak is characterised by the following 
    five parameters: amplitude-related parameter $\Sigma_\mathrm{p\odot}$, mean age 
    $\tau_\mathrm{p}$, age dispersion $d\tau_\mathrm{p}$, mean radius $R_\mathrm{p}$,
    and radial dispersion $dR_\mathrm{p}$. 
    
    \item \textit{IMF.} Our four-slope broken power-law IMF is defined by seven 
    parameters -- four slopes, $\alpha_0$, $\alpha_1$, $\alpha_2$, and $\alpha_3$, 
    and three break-point masses, $m_0$, $m_1$, and $m_2$.  
    In order to work with a different IMF, one can introduce another 
    set of IMF parameters in the \texttt{jjmodel} code.
    
    \item \textit{Metallicities.} There are four parameters describing a universal 
    thick-disk AMR -- initial and present-day metallicity,
    $\mathrm{[Fe/H]}_\mathrm{t0}$ and $\mathrm{[Fe/H]}_\mathrm{tp}$, and parameters 
    $t_0$ and $r_\mathrm{t}$ controlling the AMR shape. In the case of 
    the thin disk, there are two alternative parameter sets.
    If the thin-disk AMR is given by \mbox{Eqs. (21) and (22)} 
    in \citetalias{sysoliatina21}, there are four parameters of the local AMR: 
    $\mathrm{[Fe/H]}_\mathrm{d0}$, $\mathrm{[Fe/H]}_\mathrm{dp}$, $r_\mathrm{d}$, 
    and $q$. Power-law slopes $k_\mathrm{[Fe/H]_{d0}}$, $k_\mathrm{[Fe/H]_{dp}}$,
    $k_\mathrm{r_d}$, and $k_\mathrm{q}$ then prescribe the radial change 
    of the thin-disk AMR. When the thin-disk AMR is given by 
    \mbox{Eqs. (\ref{eq:amr_fit}) and (\ref{eq:f_amrd})}, an alternative set of parameters must be given (first raw in \mbox{Table \ref{tab:amr_params}}; 
    see \mbox{Section \ref{sect:amr_fit}} for details). 
    Parameter $d\mathrm{[Fe/H]_{dt}}$ specifies 
    the metallicity dispersion applied to the disks' mono-age sub-populations. 
    The halo MD is described by the mean metallicity 
    $\mathrm{[Fe/H]}_\mathrm{sh}$ and dispersion $d\mathrm{[Fe/H]}_\mathrm{sh}$. 
    
    \item \textit{$W$-velocities.} The thin-disk local kinematics is given by 
    the AVR power index $\alpha$ and scaling factor $\sigma_\mathrm{e}$. 
    Special kinematics of the sub-populations associated with each of the additional SFR peaks is described by $W$-velocity dispersion $\sigma_\mathrm{p}$. 
    The local dispersions of the thick disk, DM, and stellar halo are given by parameters $\sigma_\mathrm{t}$, $\sigma_\mathrm{dh}$, and $\sigma_\mathrm{sh}$, respectively. 
\end{itemize}
\fi

\subsubsection{Building the disk}\label{sect:build_disk}

To construct the MW disk in the framework of \mbox{the JJ model}, we do 
the following sequence of steps. 

To start, we specify \mbox{the JJ-model} parameters
(\mbox{Section \ref{sect:model_params}}, also 
\mbox{Tables \ref{tab:jjparams} and \ref{tab:amr_params}}).

Then, we calculate the input quantities, such 
as the radial surface density profiles of the MW components, IMF, 
local AVR, thin- and thick-disk AMR and SFR functions. 
At this step, we also optimise the DM scaling parameter $a_\mathrm{h}$ 
using the predicted circular speed at the solar radius. 

After that, we proceed with the iterative solving of the Poisson equation 
at the solar radius. The solving procedure can be described in terms of 
% A&A version
\ifx
three sub-steps.

First, the scale heights of isothermal sub-populations $h_\mathrm{j}$ 
are initialised with realistic values (\mbox{100 pc} for the gas components, 
\mbox{400 pc} and \mbox{1 kpc} for the thin and thick disk, respectively, 
and \mbox{2 kpc} for the DM and stellar halo). Additionally, we assume 
\mbox{$\sigma_\mathrm{H_2} = 3$ km s$^{-1}$} and 
\mbox{$\sigma_\mathrm{HI} = 10$ km s$^{-1}$} for the 
vertical velocity dispersion of the molecular and atomic gas. 

Second, we solve \mbox{Eq. (\ref{eq:zPhi})} with substitution 
of \mbox{Eqs. (\ref{eq:rhoz_expPhi}) and (\ref{eq:rho_j0})} up to the 
maximum value of potential, $\Phi_\mathrm{max}$, 
which is predetermined empirically at each $R$. The obtained dependence $z(\Phi)$ 
is then numerically inverted, and the result is fitted with a polynomial 
to get a smooth function $\Phi(z)$.

Third, using the derived vertical potential, we update all scale heights 
by integrating the normalised density profiles up to infinite $z$ 
(\mbox{Eq. (8)} in \citetalias{just10}). In the case of the gas components, 
we also compare their derived scale heights to the shale heights 
from \citet{nakanishi16}, which we aim to reproduce; 
the calculated difference is used as a weight for modifying 
$\sigma_\mathrm{H_2}$ and $\sigma_\mathrm{HI}$. 
Then the solving of the Poisson equation at the previous step is repeated
with the updated scale heights and gas velocity dispersions as an input. 
The iterations continue until all differences between the old and new 
scale heights, as well as between the adopted and calculated scale heights 
of the molecular and atomic gas, become less than the vertical resolution 
of the model. With \mbox{$dz=2$ pc}, 
the procedure converges after \mbox{$\sim$5} iterations. 
\fi
% Arxiv version
the following three sub-steps.
\begin{itemize}
\setlength\itemsep{0.25em} 
    \item The scale heights of isothermal sub-populations $h_\mathrm{j}$ 
    are initialised with realistic values (\mbox{100 pc} for the gas components, 
    \mbox{400 pc} and \mbox{1 kpc} for the thin and thick disk, respectively, 
    and \mbox{2 kpc} for the DM and stellar halo). Additionally, we assume 
    \mbox{$\sigma_\mathrm{H_2} = 3$ km s$^{-1}$} and 
    \mbox{$\sigma_\mathrm{HI} = 10$ km s$^{-1}$} for the 
    vertical velocity dispersion of the molecular and atomic gas. 
    
    \item We solve \mbox{Eq. (\ref{eq:zPhi})} with substitution 
    of \mbox{Eqs. (\ref{eq:rhoz_expPhi}) and (\ref{eq:rho_j0})} up to the 
    maximum value of potential, $\Phi_\mathrm{max}$, 
    which is predetermined empirically at each $R$. The obtained dependence $z(\Phi)$ 
    is then numerically inverted, and the result is fitted with a polynomial 
    to get a smooth function $\Phi(z)$.
    
    \item Using the derived vertical potential, we update all scale heights 
    by integrating the normalised density profiles up to infinite $z$ 
    (\mbox{Eq. (8)} in \citetalias{just10}). In the case of the gas components, 
    we also compare their derived scale heights to the shale heights 
    from \citet{nakanishi16}, which we aim to reproduce; 
    the calculated difference is used as a weight for modifying 
    $\sigma_\mathrm{H_2}$ and $\sigma_\mathrm{HI}$. 
    Then the solving of the Poisson equation at the previous step is repeated
    with the updated scale heights and gas velocity dispersions as an input. 
    The iterations continue until all differences between the old and new 
    scale heights, as well as between the adopted and calculated scale heights 
    of the molecular and atomic gas, become less than the vertical resolution 
    of the model. With \mbox{$dz=2$ pc}, 
    the procedure converges after \mbox{$\sim$5} iterations. 
\end{itemize}

Using the obtained local density profiles, we calculate the overall thin-disk half-thickness at $R_\odot$. Then we prescribe its value also at the non-solar
galactocentric distances (depending on the model parameters, it can be 
a constant or flaring). Similarly, we fix the half-thickness of the thick disk 
at other radii by setting it everywhere to its local value. 

At the next step, we solve the vertical Poisson equation at other radii. 
At $R \neq R_\odot$, the solving procedure is applied with some modifications. 
Together with the gas velocity dispersions, we also assume the AVR scaling parameter 
$\sigma_\mathrm{e}$ and the thick-disk velocity dispersion $\sigma_\mathrm{t}$. 
We also check the consistency of the derived thin- and thick-disk
half-thickness with their prescribed values for this radius, 
and use the discrepancy values to modify $\sigma_\mathrm{e}$ and 
$\sigma_\mathrm{t}$. 
For the iterations to stop, also the values of these discrepancies must become 
less than the vertical resolution $dz$.

All subsequent calculations can be classified as post-processing of 
the obtained self-consistent vertical potential and density profiles. 
For example, we can calculate the vertical and radial density profiles of the mono-age 
disk sub-populations, which with the help of the AMR can be transformed 
into the profiles of the mono-abundance populations. We can study 
age, metallicity, or velocity distributions at the different $R$ and $z$. 
Finally, IMF and isochrones can be used to create mock samples comparable to 
the real observable MW populations (\mbox{Section \ref{sect:sps}}).

\section{Results}\label{sect:model_predictions}

To illustrate the predictive capability of the generalised \mbox{JJ model}, 
we present its test realisation, which is consistent with 
the local calibration  from \citetalias{sysoliatina21}. Here we must note 
that not all of   
the parameters defining the MW radial structure are calibrated 
against the data or are adopted 
from the literature; for several parameters we only assume plausible 
radial trends, which need to be tested with some observational sample 
in the future. 

\subsection{Test model}\label{sect:model1}

\begin{figure*}[t]
\includegraphics[scale=0.55]{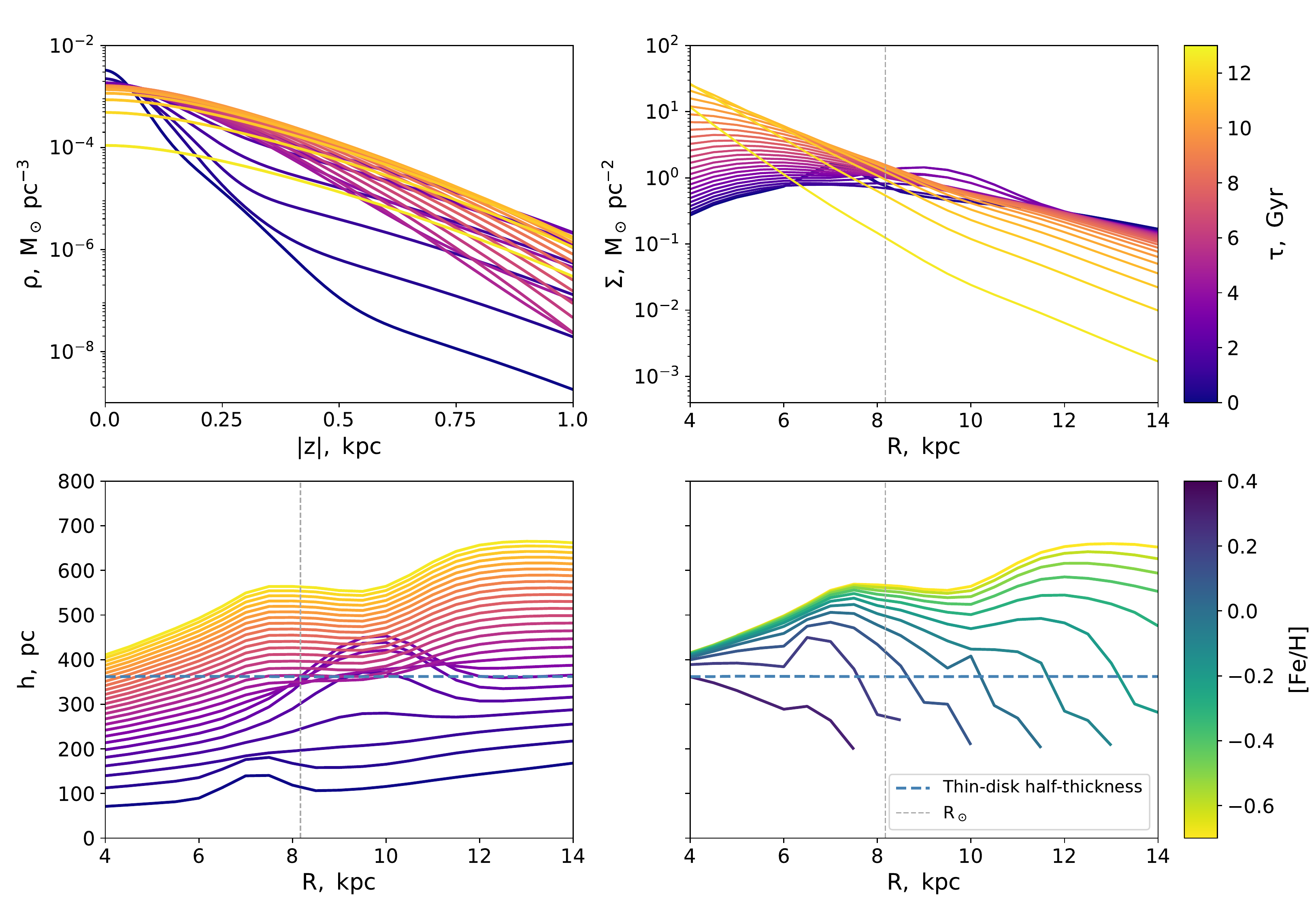}
\caption{Radial and vertical structure of the thin disk according to the 
JJ model realisation $\mathrm{TM1}$ (Section \ref{sect:model1}). 
\textit{Top.} Vertical and radial density profiles 
of the thin-disk mono-age sub-populations. 
The vertical density profiles correspond to the 
solar radius. \textit{Bottom.} Radial variation in scale heights 
of the thin-disk mono-age and mono-abundance sub-populations.}
\label{fig:rho_h}
\end{figure*}

For the test model ($\mathrm{TM1}$ hereafter), 
we adopt the following parameters (\mbox{Tables \ref{tab:jjparams} and \ref{tab:amr_params}}). 

The solar galactocentric distance is set to \mbox{$\mathrm{R_\odot = 8.17}$ kpc}
\citep{gillessen19}, and its height above the Galactic plane is 
\mbox{$\mathrm{z_\odot = 20}$ pc} (consistent with \citealp{bennett19}). 
For the solar peculiar velocity we take the classical value from 
\citet{schoenrich10}, \mbox{$V_\odot = 12.25$ km s$^{-1}$}. 

Vertically we go up to \mbox{2 kpc} with a fine \mbox{2-pc} resolution. 
Radially, we consider galactocentric distances \mbox{4--14 kpc} with 
\mbox{0.5-kpc} bin width. This large radial step is sufficient for the illustration purposes
of this section, but there are no physical or technical limitations for $dR$: 
it can be chosen to be as fine as required by the sample's modelling.

We model the thin disk of a constant thickness. 
The thin- and thick-disk scale lengths are set to \mbox{2.5 kpc} and 
\mbox{2.0 kpc}, respectively, 
which is consistent with the different observational results summarised in \citet{bhawthorn16}. 
Our choice of scale lengths of two gas components is described in 
\mbox{Section \ref{sect:gas}}. The radial surface density profiles of both 
molecular and atomic gas have a cut-off at 
\mbox{$R_\mathrm{0,H_2} = R_\mathrm{0,HI} = 4$ kpc}. 
For the slope of the stellar halo density profile we adopt 
$\alpha_\mathrm{in}=-2.5$ \citep{bhawthorn16}, 
and the bulge mass is set to \mbox{$M_\mathrm{b} = 0.8 \cdot 10^{10}$ M$_\odot$}. 
Our bulge component has a relatively small mass, because part of the b/p bulge connected 
to the bar is already included in the disk components.

The local surface densities, IMF, kinematic, thick-disk and local thin-disk SFR parameters 
are adopted from \citetalias{sysoliatina21}. 
Parameters $k_\mathrm{td_2}$, $k_\mathrm{\zeta}$, and $k_\mathrm{\eta}$, 
which prescribe the radial variation of the thin-disk SFR continuum,  
are chosen such that the inside-out disk growth is mimicked: the SF peak shifts from 
older to younger ages as we go from the inner to outer disk (upper panel of 
\mbox{Figure \ref{fig:nsfr-avr}}). 
Two extra Gaussian peaks at ages \mbox{0.5 Gyr} and \mbox{3 Gyr} are assumed to be centred 
at the radii of \mbox{7 kpc} and \mbox{9 kpc} with the radial dispersions of 
\mbox{0.5 kpc} and \mbox{1 kpc}, respectively. 

The halo MD is characterised by two parameters, 
$(\mathrm{[Fe/H]}_\mathrm{sh},d\mathrm{[Fe/H]}_\mathrm{sh}) = (-1.5,0.4)$ 
(consistent with \citealp{zuo17}). Finally, we adopt a new treatment of the thin-disk AMR 
based on the model calibration against the APOGEE RC sample (\mbox{Section \ref{sect:amr}}).
The thin- and thick-disk AMR parameters adopted for $\mathrm{TM1}$
are listed in \mbox{Table \ref{tab:amr_params}}. We assume no scatter 
in metallicity for the disk sub-populations.

\subsection{Predictions}\label{sect:predictions}

We concentrate on the following model predictions: vertical and radial 
density profiles, scale heights of the mono-age and mono-abundance thin-disk
populations, and $W$-velocity distribution functions of several well-defined 
mock populations, such as G dwarfs and RC stars.

\subsubsection{Radial and vertical density profiles}\label{sect:profiles}

The top panel of \mbox{Figure \ref{fig:rho_h}} shows the vertical and radial
thin-disk density profiles predicted for \mbox{0.25-Gyr} age bins. 
The vertical profiles are calculated for the solar radius. 
The radial profiles correspond to the midplane densities.

The vertical density profiles for ages \mbox{$\gtrsim4$ Gyr}  
demonstrate a monotonous decline with $z$. Though the youngest age bins 
are characterised by a more complex vertical density fall-off, 
we can identify several regimes with the different slopes. This behaviour is 
a direct consequence of the model input. In $\mathrm{TM1}$, the mono-age
sub-population fractions associated with the extra SFR peaks are more 
dynamically heated than their remaining parts following AVR 
(see the bottom panel of \mbox{Figure \ref{fig:nsfr-avr}}). Thus, the youngest 
age bins consist of several kinematically different sub-components, and the final 
shape of the vertical profile depends on these sub-components' fractions and 
$W$-velocity dispersions. We observed this multi-slope shape of the vertical 
density profile for the local A-type star sample selected from 
\mbox{the \textit{Gaia} DR2} (\citetalias{sysoliatina21}). 

As to the radial profiles, we see that though the overall thin-disk surface 
density law is assumed to be exponential (\mbox{Eq. (\ref{eq:disk_sigma})} 
and \mbox{Figure \ref{fig:rhor}}), the profiles of individual age bins can 
significantly deviate from this shape. The oldest thin-disk populations have  
quasi-exponential radial profiles, though two-slope exponential law can 
provide a more accurate fit in this case. Starting from \mbox{$\sim$8--9 Gyr},
profiles have a peak at the radius, which shifts to the outer disk as the age
decreases. This reflects the lack of young populations in the inner disk due 
to the chosen SFR: young thin-disk stars formed mostly at large radii.
The density profiles of the youngest populations have additional overdensities 
located around the extra SFR peaks' mean radii $R_\mathrm{p}$. 

This result is in a good agreement with the observed radial surface density trends 
in the MW disk. \citet{bovy16} and \citet{mackereth17} used APOGEE data to study 
the radial disk structure in terms of mono-age and mono-abundance populations. 
They find that the radial surface density profiles of both mono-age and mono-abundance 
low-$\alpha$ populations (which correspond to the chemically defined thin disk) 
are at best described by a broken exponential law. 
High-$\alpha$ populations representing the thick disk show no significant change 
of the slope over the studied distance range of \mbox{3--15 kpc}. 
Similar radial structure 
follows from the JJ-model predictions (see the upper-right panel of \mbox{Figure \ref{fig:rhor}}, 
which shows radial surface density profiles of the mono-age thin-disk sub-populations). 
As to the thick disk in the JJ model, its mono-age and mono-abundance sub-populations 
are all single exponential profiles with a slope of the overall thick-disk radial profile 
shown in \mbox{Figure \ref{fig:rhor}}. \citet{amores17} used BGM in combination with 
the 2MASS data to study the disk properties such as warp, flare, and populations' scale lengths.
They find that the thin-disk scale length is age dependent, with the youngest populations
characterised by the largest scale lengths. The JJ model also recreates this trend which 
naturally follows from the inside-out disk growth scenario implied by the shape of our adopted SFR 
(\mbox{Figure \ref{fig:nsfr-avr}}). The same negative correlation of scale length with age and 
double-slope radial profiles of the thin-disk mono-age sub-populations are predicted by hydrodynamic
simulations of the MW-like galaxies in \citet{minchev15}.

\subsubsection{Scale heights}\label{sect:scaleheights}

The lower panel of \mbox{Figure \ref{fig:rho_h}} displays the radial change 
of the thin-disk scale heights calculated for different age and metallicity bins. 

One of the interesting results of our modelling is that the mono-age populations 
flare even though the overall thickness of the thin disk is the same 
at all radii (dashed blue line in the lower-left panel of 
\mbox{Figure \ref{fig:rho_h}}). 
The same effect of mono-age sub-populations flaring when the overall disk 
thickness is approximately constant
is reproduced in \citet{minchev15}. The effect is generated by an interplay of two factors: 
(1) the inner disk is on average more dynamically heated than the outer disk, and (2) 
the relative fraction of old-to-young stars decreases with increasing radius. 
Similarly, \citet{mackereth17} find flaring for essentially all mono-age sub-populations 
of the low-$\alpha$ APOGEE stars.
As follows from \mbox{Figure \ref{fig:rho_h}}, for the youngest and oldest age bins 
in the range of \mbox{4--14 kpc} scale heights change from approximately 
\mbox{80 pc} to \mbox{180 pc} and from \mbox{440 pc} to \mbox{700 pc}, respectively.  
Thus, flaring scale lengths predicted by this model lay in the range 
of \mbox{$\sim$12.5--17 kpc}. 
However, \citet{mackereth17} report a trend for the flaring with age
that is the oppositive of that given by the JJ model: according to the APOGEE data, 
the youngest populations flare the most. \citet{mackereth17} 
suggest that at the early stage of the disk formation flaring could be suppressed 
by the active accretion. This topic may be deeper investigated in our future works.
Another feature is the presence of a peak in the scale-height profiles of the 
young populations due to the contribution of kinematically hot SFR peaks'
sub-populations. Also, we see an opposite `response'
in the scale heights of older ages around \mbox{9 kpc}: here scale heights 
grow slower than we would expect from the trends up to this radius. We can attribute
this behaviour to the SFR re-normalisation. As explained in 
\mbox{Section \ref{sect:thindisk_sfr}}, we re-normalise the total thin-disk 
SFR to the same $\Sigma_\mathrm{d}$ after adding extra peaks. In the case 
of $\mathrm{TM1}$ we have two extra peaks at young ages, that is, we increase 
the contribution of the young sub-populations by decreasing contributions 
of all other ages. And though these young sub-populations are partly more heated than prescribed 
by AVR, the modified age balance may in some age bins result in a more 
plane-concentrated matter distribution than we could expect without extra SFR peaks. 

As mono-age populations are not easily observable, it is useful to convert 
them to mono-abundance populations using the AMR function 
(\mbox{Section \ref{sect:amr_fit}}). At the lower right panel of 
\mbox{Figure \ref{fig:rho_h}} we show the radial variation of the 
thin-disk scale heights calculated in 10 metallicity bins on the interval 
from $-0.7$ to \mbox{0.3 dex} with a \mbox{0.1-dex} bin width. 
We see that the scale-height radial profiles of the metal-poor populations 
closely follow the corresponding profiles of the oldest age bins. 
In other words, mono-abundance thin-disk populations can be also considered 
as mono-age when the metallicity is low. 
To the same conclusion arrived \citet{minchev17} who investigated the relationship 
between mono-age and mono-abundance sub-populations using the chemodynamical model 
of the MW-like galaxy with inside-out disk growth. As the metallicity increases, 
mono-abundance populations begin to include more ages, and the corresponding
age distributions depend on radius. For metallicities \mbox{$>-0.2$ dex}, 
profiles do not span over the whole range of galactocentric distances, but 
end at some radius, and this stop point moves to the inner disk as the 
metallicity increases. This happens because a given metallicity ceases 
to be present starting from some radius, according to our simple chemical 
evolution model. Due to the multi-age composition of most of the mono-abundance populations, 
the flaring trend is observed only for most metal-poor bins, while for other 
mono-abundance sub-populations it is suppressed. Also, in \citet{minchev17} the authors 
find that flaring is not prominent in  
most of the metallicity bins because of their negative radial age gradient. On the other side, 
radial surface density profiles of APOGEE mono-abundance sub-populations show flaring 
in almost all metallicity bins \citep{bovy16,mackereth17}. This can be viewed as a lack of  
flaring in the overall thin-disk profile, which can be easily included into the model. 
Also, this can point to the lack of some fundamental physical process, such as 
radial migration, that would relocate some fraction of stars from the inner to the outer disk 
and thus lead to an extra flaring.

\begin{figure}[t]
\includegraphics[width=\columnwidth]{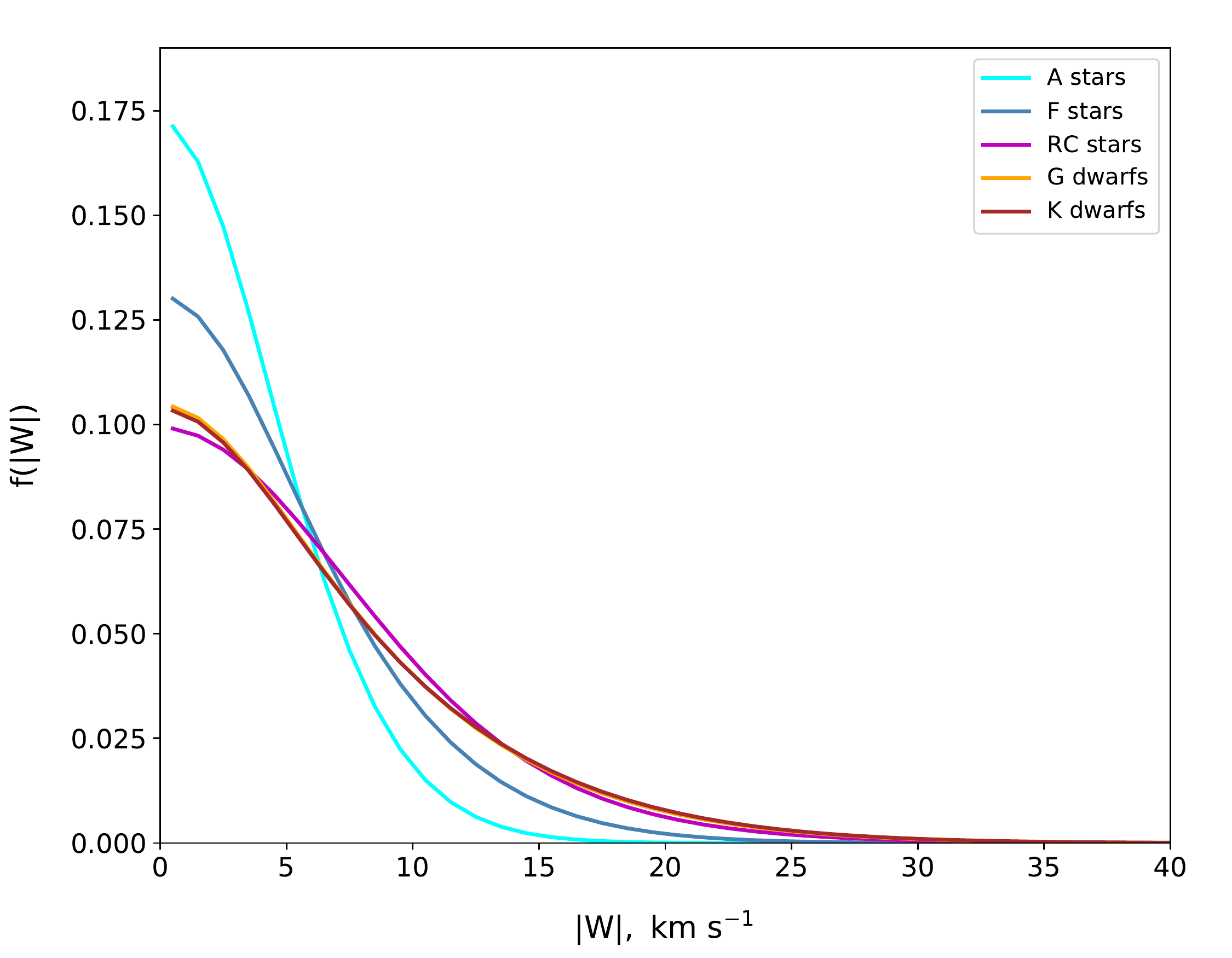}
\caption{Normalised $W$-velocity distribution functions of the different 
mock populations at the solar radius.}
\label{fig:fW}
\end{figure}

As to the thick disk, our non-flaring disk is consistent with the observed nearly constant
thickness of the high-$\alpha$ APOGEE mono-abundance sub-populations \citep{bovy16,mackereth17},
properties of the thick disk constrained with the SDSS data in the BGM framework \citep{robin14}, 
and the thick disk simulated in \citet{minchev15}.

\subsubsection{Vertical kinematics}\label{sect:kinematics}

Following a routine described in \mbox{Section \ref{sect:sps}}, we converted 
the modelled densities into SA using PARSEC isochrones. Then we created several 
mock samples of the well-defined observable populations, such as A, F, and RC stars 
and G and K dwarfs. All of them, except RC stars, were selected by simple cuts applied 
to the effective temperature and surface gravity (in the case of dwarfs). 
Our RC modelling routine is described in \mbox{Section \ref{sect:RC_in_model}}. 

\mbox{Figure \ref{fig:fW}} shows $W$-velocity distribution functions of these five mock 
thin-disk samples calculated for the solar radius for \mbox{$|z|< 2$ kpc}. 
As expected, A- and F-star samples 
consisting exclusively or predominantly from the young populations, have $f(|W|)$ 
with a prominent core. The main sequence (MS) stars, G and K dwarfs, have velocity 
distributions with a much more significant contribution of the dynamically heated stars. 
Interestingly, the RC sample consisting of a mixture of different ages with a significant 
fraction of young populations, has $f(|W|)$ essentially indistinguishable from the one 
of the MS samples (also see Fig. 8 and 10 in \citetalias{sysoliatina21}). 
For the current model realisation, 
this can be understood as follows: though the RC sample contains more 
young stars than G and K dwarfs, young populations of the SFR peaks
also partly contribute to the tail of the velocity distribution, such that the resulting 
RC and G- and K-dwarf $f(|W|)$ are very similar. \mbox{Figure \ref{fig:fW}} 
is also representative for other modelled radii.

\section{AMR from APOGEE RC stars}\label{sect:amr}

In this section, we take the first step towards calibration of  
the generalised \mbox{JJ model} across the disk. We use the developed machinery 
to constrain the AMR function with the APOGEE data covering distances 
\mbox{4 kpc $< R < $ 14 kpc}. Motivated by the obtained results, we then 
develop a new analytic representation of the thin-disk AMR.  

\subsection{Sample selection}\label{sect:apogee_data}

To constrain the AMR across a wide range of galactocentric distances, 
we use the latest data release, DR16, of the APOGEE RC catalogue 
\citep{bovy14,ahumada20}.
The APOGEE RC catalogue comprises a kinematically unbiased sample of 39\,675
bright stars spanning over a large fraction of the Galactic disk. 
As the luminosity of RC stars depends very little on their chemical composition and age, 
they can play the role of standard candles. Thus, for the RC stars, 
it is possible to obtain robust photometric distances -- 
distance uncertainties of the APOGEE RC stars 
are reported to be less than $\sim$5\% \citep{bovy14}.
Using heliocentric distances provided in the RC catalogue, we calculated 
Galactic cylindrical coordinates $R$ and $z$ 
(the adopted position of the Sun $(R_\odot,z_\odot)$ is the same as in 
\mbox{Table \ref{tab:jjparams}}). 

\begin{figure}[t]
\subfloat{\includegraphics[width=\columnwidth]{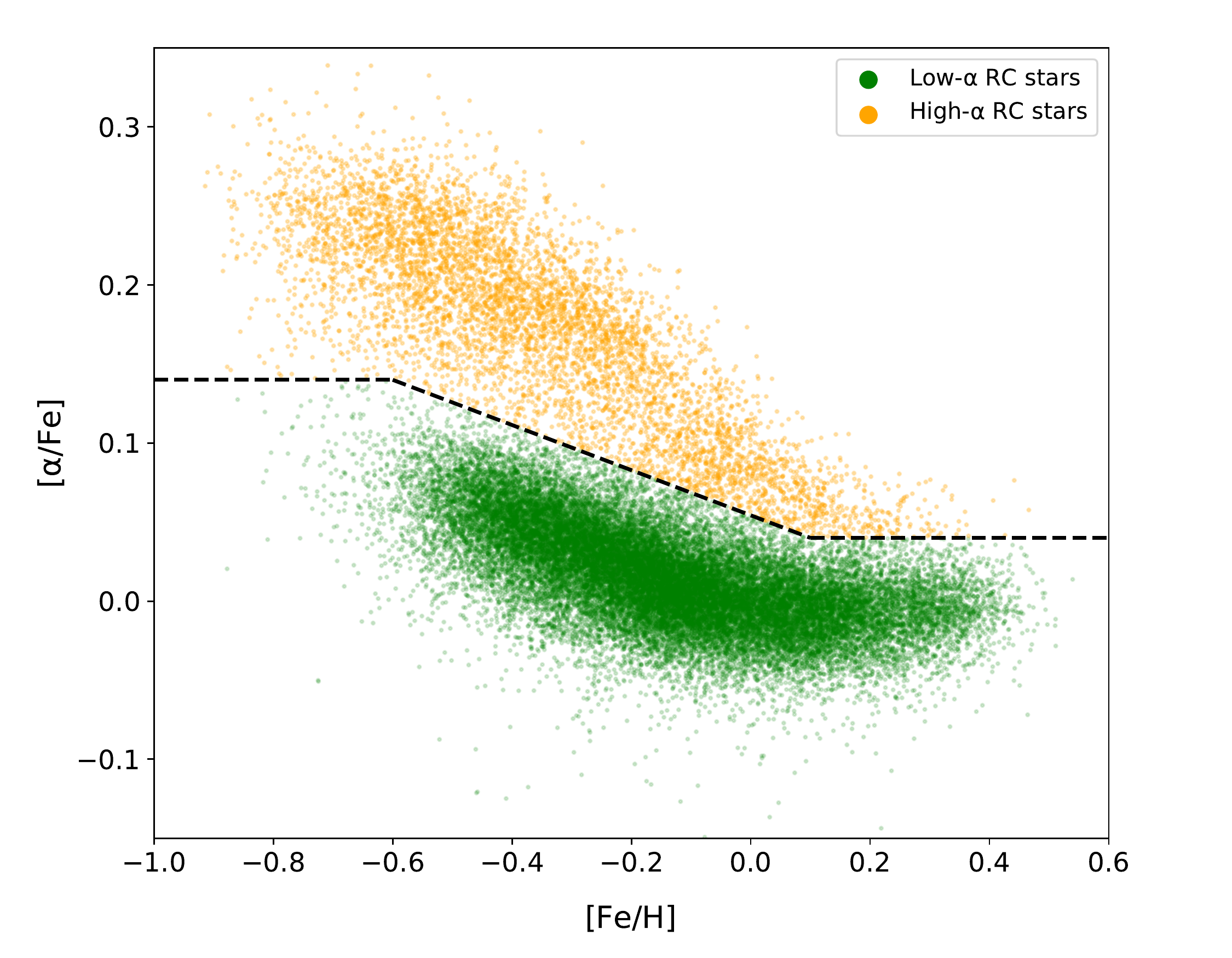}} 
%width=\columnwidth % scale=0.55 - for referee

\subfloat{\includegraphics[width=\columnwidth]{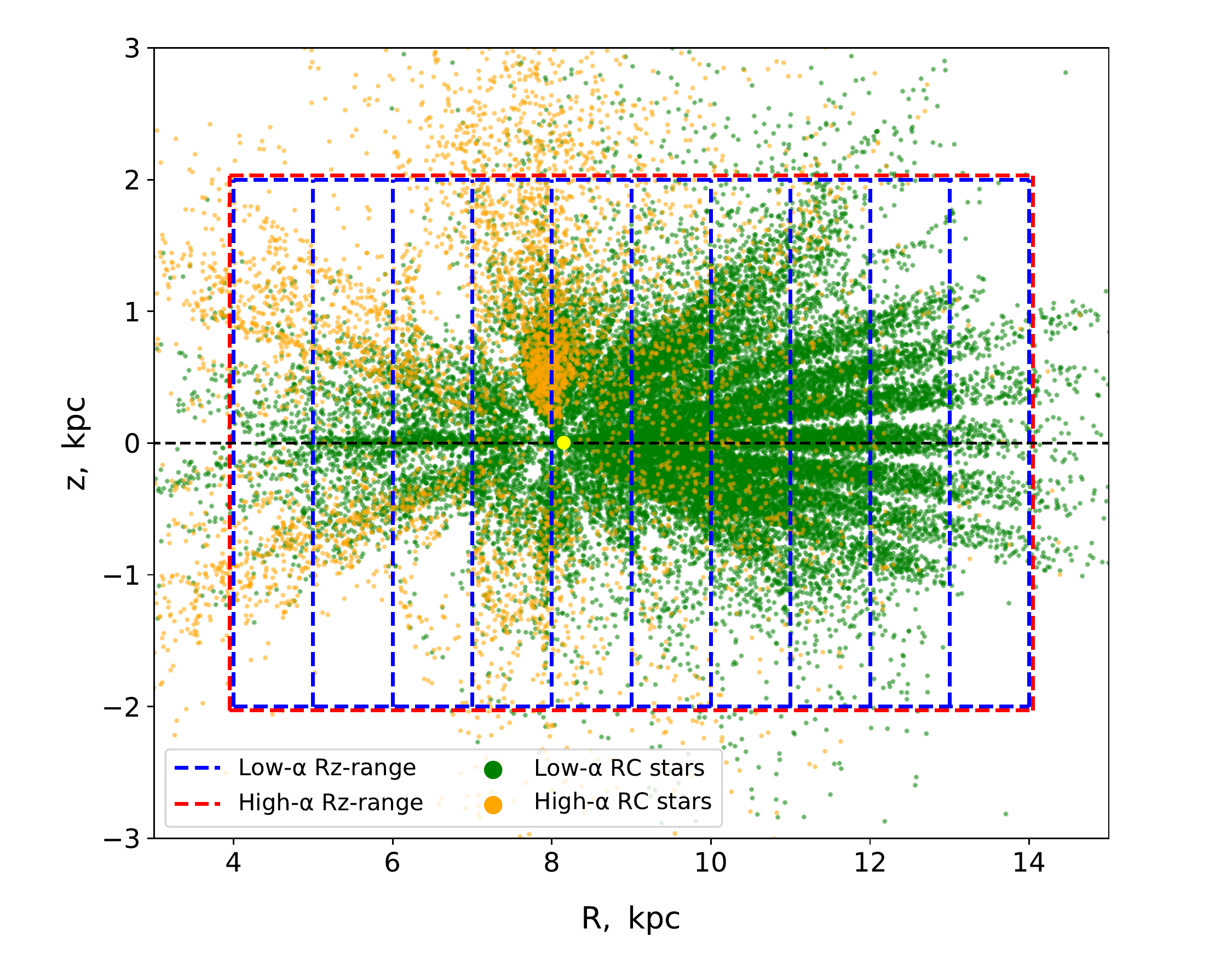}}
\caption{Selection of the APOGEE RC samples. 
\textit{Top}. Low-$\mathrm{\alpha}$ (green) and high-$\mathrm{\alpha}$ (orange) 
RC sequences. The separation border corresponding to \mbox{Eq. (\ref{eq:chem_disks})} 
is shown with dashed black line.
\textit{Bottom}. Selected low-$\mathrm{\alpha}$ and high-$\mathrm{\alpha}$ 
RC populations in the $R$-$z$ plane. 
Dashed lines mark the adopted limits for the galactocentric distance and height.
The radial binning of the low-$\mathrm{\alpha}$ population corresponds to $\mathrm{run0}$ 
(see text). The adopted position of the Sun is marked with a yellow dot.}
\label{fig:rc_data}
\end{figure}

It is well known that when projected onto [$\mathrm{\alpha/Fe}$]-[$\mathrm{Fe/H}$] 
chemical abundance plane, the MW disk stars form two distinct sequences, which are  
associated with the thin- and thick-disk populations. 
We used a simple criterion to separate low- and \mbox{high-$\mathrm{\alpha}$} sequences, similar to the one 
adopted in \mbox{\citetalias{sysoliatina21}}\footnote{In our previous work,
we used an older release of the APOGEE RC catalogue, DR14, so the given coefficients 
of \mbox{Eq. (\ref{eq:chem_disks})} do not exactly match the corresponding coefficients 
in \mbox{\citetalias{sysoliatina21}} (Eq. 6), this is related to 
the difference in DR14 and DR16 calibrations.} 
\mbox{(Figure \ref{fig:rc_data}, top panel)}: 
\begin{equation}
\mathrm{[\alpha/Fe]} = 
  \begin{cases}
      \ 0.04, & \mathrm{[Fe/H]} > 0.1 \\ 
      \ -0.14  \, \mathrm{[Fe/H]} + 0.05, & -0.6 < \mathrm{[Fe/H]} < 0.1 \\
      \ 0.14, & \mathrm{[Fe/H]} < -0.6. 
  \end{cases}
\label{eq:chem_disks}
\end{equation}
To eliminate halo stars, we applied a cut $\mathrm{[Fe/H]} > - 1$. 
From the remaining subsample, we took stars with $\mathrm{|z|} < 2 $ kpc, both for 
the high- and \mbox{low-$\mathrm{\alpha}$} population. 
Finally, we defined the radial range of the samples. 
Here we keep in mind that the \mbox{high-$\mathrm{\alpha}$} population is known 
to be chemically homogeneous, while the \mbox{low-$\alpha$} MD 
systematically change with galactocentric distance. 
Therefore, in case of  \mbox{high-$\mathrm{\alpha}$} sample,
we selected stars with galactocentric distances \mbox{4 kpc $< \mathrm{R} <$ 14 kpc} and did not
apply radial binning (red dashed box at the lower panel of \mbox{Figure \ref{fig:rc_data}}). 
For the  \mbox{low-$\mathrm{\alpha}$} stars, we used a grid of \mbox{1-kpc} bins centred 
at \mbox{$4.5,5.5,...,13.5$ kpc} (blue dashed lines). 
We later refer to this grid as $\mathrm{run0}$. 
In order to eliminate an impact of binning on the reconstructed AMR, 
we used four additional grids 
for the  \mbox{low-$\mathrm{\alpha}$} sample (\mbox{Section \ref{sect:amr_reconstruction}}): 
three grids shifted with respect to the initial one by $+0.25$, $+0.5$, and 
 \mbox{$+0.75$ kpc}
($\mathrm{run1}$, $\mathrm{run2}$, and $\mathrm{run3}$, respectively) 
and a finer grid with  \mbox{0.5-kpc} bins centred at  \mbox{$4.25,4.75,...,13.75$ kpc} ($\mathrm{run4}$). 
Also, we found that bins of these grids that partially lay at  \mbox{$\mathrm{R} > 14$ kpc} and contain 
fewer than $\sim300$ stars cannot provide enough data to set a robust constraint on the 
thin-disk AMR parameters. For this reason, we excluded such bins from our analysis. 

In the selected samples, 95\% of stars have a signal-to-noise ratio (S/N) $\gtrsim50$,
so we did not apply any quality cuts. 
Our final RC samples contain 4\,444 ( \mbox{high-$\mathrm{\alpha}$}) and 32\,917 stars ( \mbox{low-$\mathrm{\alpha}$}, $\mathrm{run0}$). 

It is also worthwhile to address the problem of extinction, which is known to be very 
significant in the Galactic plane at the large distances reached by our sample. 
As explained in \citet{majewski11}, the APOGEE photometry is corrected for extinction 
with the help of the Rayleigh-Jeans colour excess (RJCE) method. The RJCE technique 
is based on sampling the spectral energy distribution (SED) in the Rayleigh-Jeans regime 
where the shape of stellar SED is essentially independent of its spectral type. 
Thus, extinction for each star can be robustly estimated by measuring the colour excess from 
its photometry in near- and mid-infrared bands. As the RC distances are based on 
the APOGEE photometry corrected for extinction, they are not biased. As reported 
in \citet{bovy14}, uncertainties in extinction determination make only a minor contribution
to the total error budget of the RC distances. The stellar parameters that were used to 
select the RC stars from the full APOGEE catalogue, such as metallicity, $\log{g}$, $T_\mathrm{eff}$, 
were derived with the APOGEE Stellar Parameter and Chemical Abundances Pipeline (ASPCAP). 
The effect of reddening is suppressed in ASPCAP by working with the observed and 
template spectra that are continuum-normalised \citep{garcia-perez16}. 
Taking all this into account, we can conclude that no special measures need to be taken in our analysis to account for the extinction, and we can compare the APOGEE RC sample to our mock 
RC population simulated with the stellar parameters and distances not affected by extinction.

\subsection{AMR reconstruction}\label{sect:amr_reconstruction}

Our approach for the AMR reconstruction is based on a direct comparison of the 
observed MDs to the modelled age distributions of some population 
and was explained in detail in \citetalias{sysoliatina21}. 
Here we summarise its main steps as applied to this study.
% A&A version
\ifx
First, we calculated the observed 
cumulative metallicity distribution functions (CMDFs) of 
the low- and \mbox{high-$\alpha$} stars using the selected RC data samples. 
In the case of the \mbox{low-$\alpha$} population, CMDF was obtained separately 
in the different radial bins. 
    
Then assuming AMR for the thick disk (same at all radii)  
and for the thin disk (radially dependent) and working in the framework 
of the generalised \mbox{JJ model}, we derived the total vertical 
gravitational potential and 
the vertical density profile of the RC population at each radius. 
The density profile was then used 
to calculate the cumulative age distribution functions (CADFs) of the thin- and 
thick-disk RC stars within the selected $R$-$z$ limits. 
    
At the next step we compared the observed CMDFs to the predicted CADFs: we 
put in correspondence stellar metallicities and ages, and thus, arrived at the 
updated AMR of the thin- and thick-disk  
(radially dependent and the same for all $R$, respectively). 
To achieve self-consistency, we iterated over the last two steps 
until the obtained thin- and thick-disk AMR converged to a constant shape 
($\sim$3--5 iterations).
    
After that we fit the obtained thin- and thick-disk AMR with analytic functions. 
In the case of the thin disk, we investigated the radial trends of the AMR best-fit 
parameters and also fit them with simple laws. As a result, we obtained 
a `generalised' thin-disk AMR fit, which is well-consistent with the originally 
reconstructed `raw' AMR. 

Finally, we performed a consistency test: 
calculated the thin- and thick-disk MDs of the 
RC stars using the updated AMR and compared these predictions to the data.
\fi
% Arxiv version

\begin{itemize}
\setlength\itemsep{0.25em} 
    \item We calculated the observed 
    cumulative metallicity distribution functions (CMDFs) of 
    the low- and \mbox{high-$\alpha$} stars using the selected RC data samples. 
    In the case of the \mbox{low-$\alpha$} population, CMDF was obtained separately 
    in the different radial bins. 
    
    \item Assuming AMR for the thick disk (same at all radii)  
    and for the thin disk (radially dependent) and working in the framework 
    of the generalised \mbox{JJ model}, we derived the total vertical 
    gravitational potential and 
    the vertical density profile of the RC population at each radius. 
    The density profile was then used 
    to calculate the cumulative age distribution functions (CADFs) of the thin- and 
    thick-disk RC stars within the selected $\mathrm{R\text{-}z}$ limits. 
    
    \item We compared the observed CMDFs to the predicted CADFs: we 
    put in correspondence stellar metallicities and ages, and thus, arrived at the 
    updated AMR of the thin- and thick-disk  
    (radially dependent and the same for all $R$, respectively). 
    To achieve self-consistency, we iterated over the last two steps 
    until the obtained thin- and thick-disk AMR converged to a constant shape 
    ($\sim$3--5 iterations).
    
    \item We fit the obtained thin- and thick-disk AMR with analytic functions. 
    In the case of the thin disk, we investigated the radial trends of the AMR best-fit 
    parameters and also fit them with simple laws. As a result, we obtained 
    a `generalised' thin-disk AMR fit, which is well-consistent with the originally 
    reconstructed `raw' AMR. 
    
    \item Finally, we performed a consistency test: 
    calculated the thin- and thick-disk MDs of the 
    RC stars using the updated AMR and compared these predictions to the data.
\end{itemize}

Each of these steps is described in more detail in the successive subsections.

\subsubsection{Cumulative metallicity distributions}\label{sect:cmdfs}

\begin{figure}[t]
\includegraphics[width=\columnwidth]{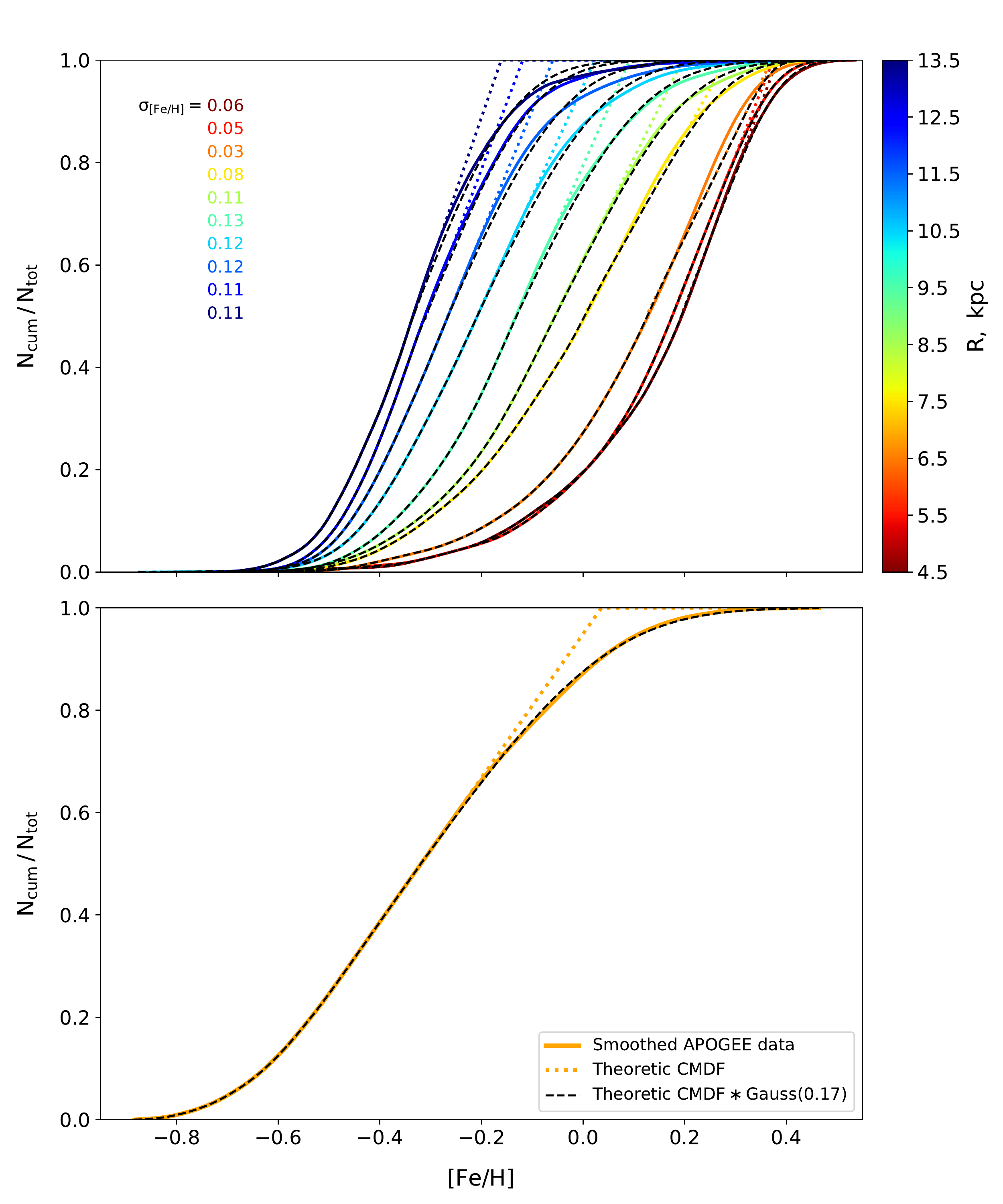}
\caption{CMDFs of the APOGEE RC samples. 
\textit{Top}. \mbox{Low-$\alpha$} CMDFs calculated in the different radial bins 
($\mathrm{run0}$ grid). Dotted coloured curves are CMDFs with extrapolated 
linear parts. Dashed black curves are $C(\mathrm{[Fe/H]})$, convolutions of the 
`deconvolved' CMDFs with the Gaussian kernel (\mbox{Eq. (\ref{eq:amr_convol})}). 
The kernel dispersions $\sigma_\mathrm{[Fe/H]}$ are shown on the plot with the same 
colour coding. \textit{Bottom}. Same as the top panel, but for the 
\mbox{high-$\alpha$} stars and no radial binning.}
\label{fig:rc_cmdf}
\end{figure}

The CMDFs of the selected low- and \mbox{high-$\alpha$} APOGEE RC stars
are shown in \mbox{Figure \ref{fig:rc_cmdf}}. 
There are three distinct regimes that can be identified for all CMDFs: 
a rapid non-linear increase at low metallicities, 
linear part, and a non-linear deceleration in growth at high metallicities.
Following \citetalias{sysoliatina21}, we make a simplifying assumption: 
at each radius, chemical enrichment process was monotonous in time (i.e. 
the AMR of interest is a monotonously growing function of Galactic time). 
This implies that low- and high-metallicity tails of CMDFs 
correspond to old and young populations, respectively.  
Then the non-linear growth of a CMDF at low metallicities   
reflects a quick chemical enrichment at the beginning of the MW disk evolution 
when the SF processes, and therefore the enrichment of ISM with metals, were most 
intense. 
The non-linear shape of CMDF at the high-metallicity end is, however, puzzling 
in this framework, as we do not expect a significant increase in 
the ISM chemical enrichment rate at the present day and do not have observational data 
that imply that. So, in order to obtain the AMR that does not show any special 
behaviour at the present day, we represented the upper part of each CMDF as 
a convolution of the linear trend (monotonous chemical enrichment) with a Gaussian kernel 
(error model). The analytic form of such a convolution was given in
\citetalias{sysoliatina21} (\mbox{Eq. 17}), and for the sake of completeness 
of narration, we show it here: 
\begin{align}
 \label{eq:amr_convol}
 C(\mathrm{[Fe/H]}) & = \frac{1}{2}\left[A\,\mathrm{Erf}\left(\frac{A}{B}\right) -  
          \left(A - 1\right)\mathrm{Erf}\left(\frac{A-1}{B}\right) + 1 \right] \nonumber \\ 
      & + \frac{B}{2 \sqrt{\pi}}\left[\exp{\left(-\frac{A^2}{B^2}\right)} -     \exp{\left(-\frac{(A-1)^2}{B^2}\right)} \right], \\
 \text{where} \quad 
 A & = k \cdot \mathrm{[Fe/H]} + b, \quad B = \sqrt{2} k \sigma_\mathrm{[Fe/H]} \nonumber
\end{align}
Parameters $k$ and $b$ are slope and intercept of the linear part of CMDF, 
and $\sigma_\mathrm{[Fe/H]}$ is a dispersion of the Gaussian kernel. 

For each CMDF, we selected the linear part and obtained parameters $k$ and $b$
($\mathrm{ 0.4 < N_{cum}/N_{tot} < 0.8}$ 
for the \mbox{low-$\alpha$} stars with bin centres at \mbox{$R < 7$ kpc} and 
$\mathrm{ 0.3 < N_{cum}/N_{tot} < 0.7}$ at all other radii and for the 
\mbox{high-$\alpha$} population).
Then, with $k$ and $b$ known, we fit the upper parts of CMDFs with \mbox{Eq. (\ref{eq:amr_convol})} and thus determined $\sigma_\mathrm{[Fe/H]}$ -- dispersion required 
to reproduce the high-metallicity CMDF tail with the Gaussian error model. The corresponding values of $\sigma_\mathrm{[Fe/H]}$ for the different $R$ bins 
of the \mbox{low-$\alpha$} population 
are displayed at the upper panel of \mbox{Figure \ref{fig:rc_cmdf}}. 
Later we used these values to model the high-metallicity MD parts. 
The `deconvolved' upper parts of CMDFs are shown with dotted lines in 
\mbox{Figure \ref{fig:rc_cmdf}}. 

The error model can represent different aspects (or be a result of their interplay):
(1) systematic errors at the metal-rich end or (2) the presence of the real stars 
that have migrated to a given volume 
from the inner radii and therefore have an imprint of the different environment 
and probably do not belong to the youngest population. 
Which of these interpretations is correct, and to which extent, 
we do not investigate within our approach.

\subsubsection{RC star modelling}\label{sect:RC_in_model}

\begin{figure*}[t]
  \subfloat{\includegraphics[scale=0.54]{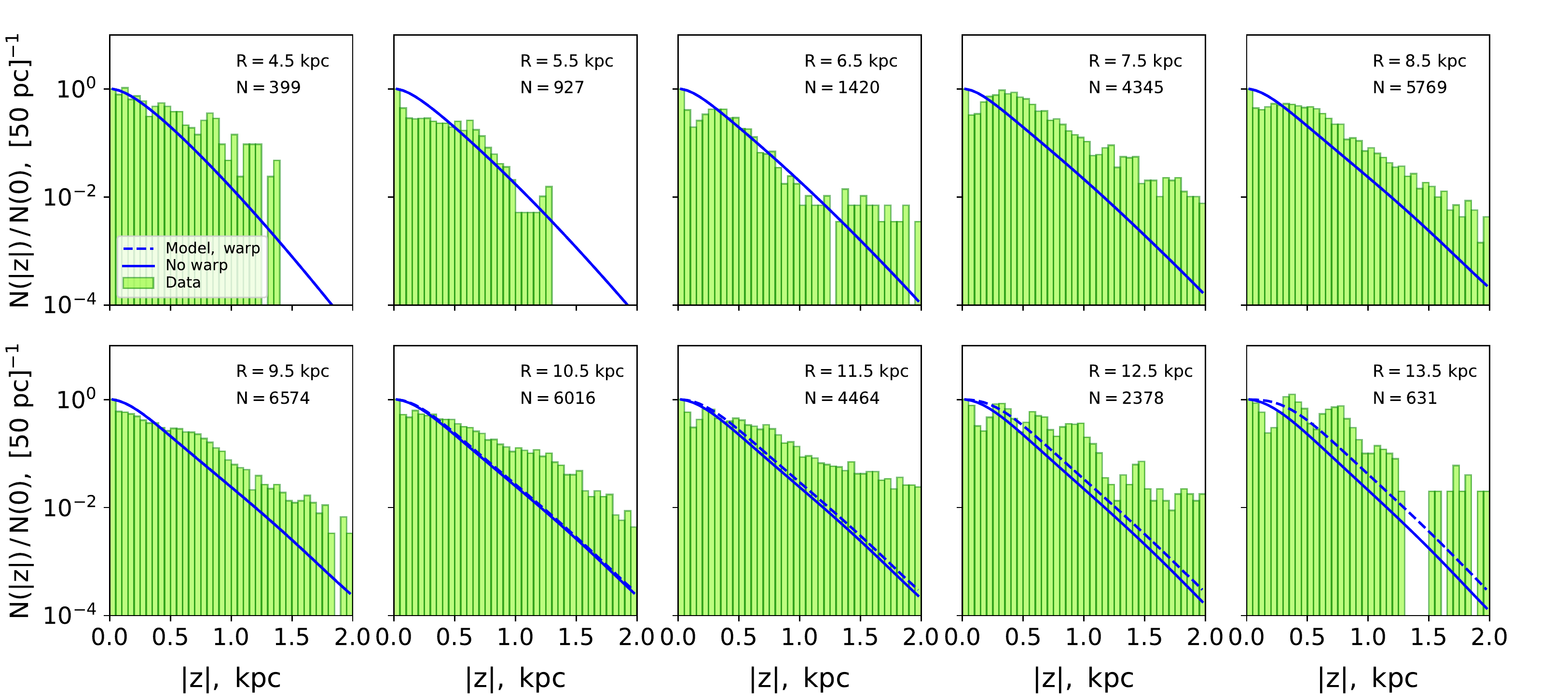}}
  
  \subfloat{\includegraphics[scale=0.54]{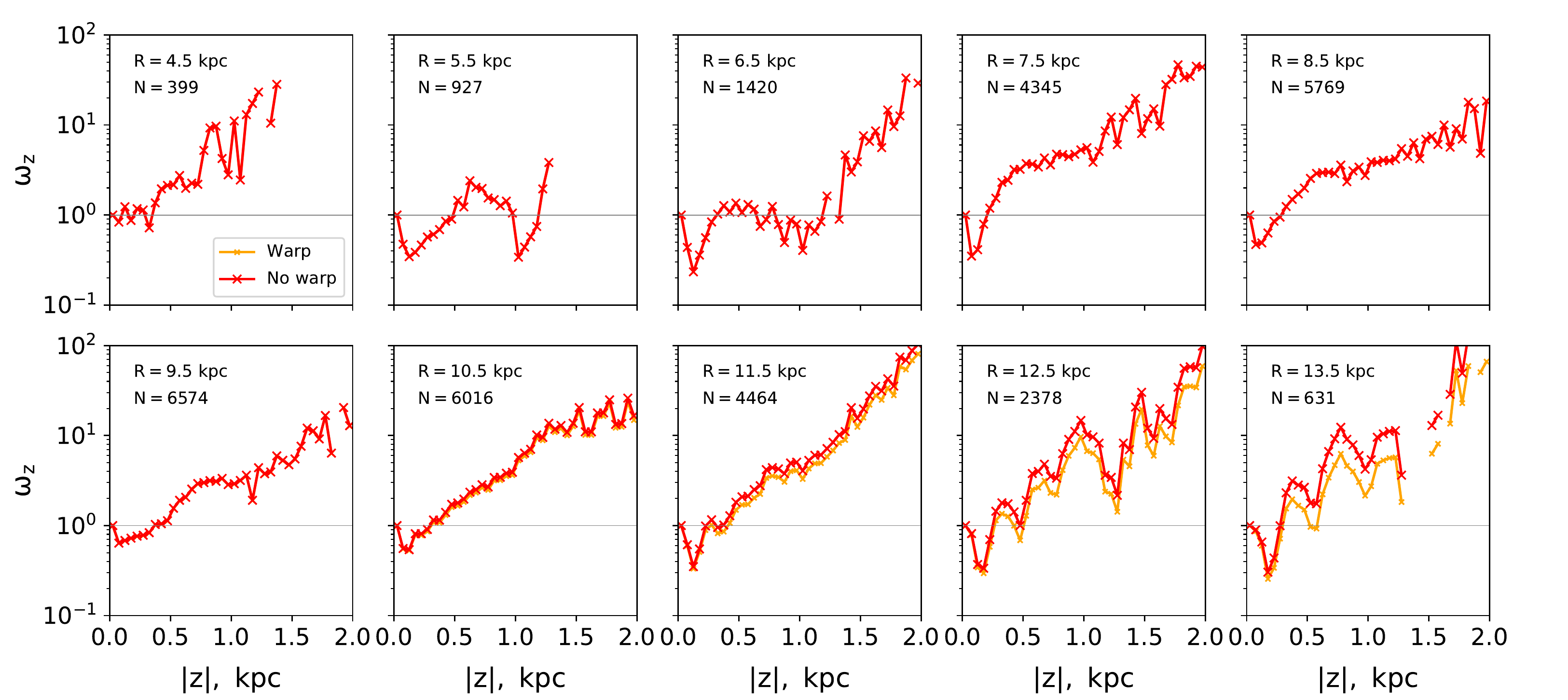}}
\caption{Spatial selection function of the thin-disk RC sample. 
\textit{Top}. Observed (green histogram) and predicted (blue curves) 
number density laws of a complete sample in $R$ bins. 
Modelled profiles with the disk warp taken into account 
are shown in dashed blue. Both observed and modelled density laws are normalised 
to the midplane densities. 
\textit{Bottom}. $z$ weights calculated as a ratio of the observed to 
modelled profiles. The grey line marks $\omega_\mathrm{z}=1$, the case of a complete sample.}
\label{fig:z_weights}
\end{figure*}

\paragraph{$\mathrm{TM0}$ framework.} For the AMR reconstruction, we used a simpler 
version of the test realisation of the generalised \mbox{JJ model} presented 
in \mbox{Section \ref{sect:model1}}. The most uncertain ingredient of $\mathrm{TM1}$ 
is the extra SFR peaks at $R \neq R_\odot$. In order not to propagate this uncertainty 
to our results and also following \citetalias{sysoliatina21}, we omitted 
the additional SFR peaks for the RC sample modelling. 
To initialise AMR at the second step of the AMR reconstruction scheme, 
we used the thin- and thick-disk AMR from \citetalias{sysoliatina21} calibrated 
against the local APOGEE RC sample (\mbox{Eqs. (21) and (22)}). 
The thin-disk AMR was extended to other radii by assuming power-law radial variations 
of its four parameters. The power indices were chosen such that the radially dependent 
AMR predicted quicker and more intense chemical enrichment in the inner disk: 
$k_\mathrm{[Fe/H]_{d0}}=-0.025$, $k_\mathrm{[Fe/H]_{dp}}=-0.0875$, 
$k_\mathrm{r_d}=0.005$, $k_\mathrm{q}=-0.03$. 
\paragraph{RC mock sample.} 
To select the RC stars in $\mathrm{TM0}$, we used the following criteria for the 
effective temperature, surface gravity, and luminosity: 
\mbox{4250 K $ < T_\mathrm{{eff}} < $  5250 K},
\mbox{$\log{g} < $ 2.75}, \mbox{1.65 $ < \log{(L/L_\odot)} < $ 1.85}. 
These cuts determine a sample of RC contaminated by red giant branch (RGB) stars. 
To remove RGB stars, \citet{bovy14} use an additional sloping cut in $\log{g}$,
a linear function of the effective temperature and metallicity:
$\log{g} < f(T_\mathrm{eff},\mathrm{[Fe/H]})$. As we have already found in 
\citetalias{sysoliatina21} (Eqs. (18)-(20)), the exact values of this cut's slope and 
intercept depend on the stellar evolution library, and also on the population's MD 
(i.e., on the assumed AMR). When visualised in
($\log{g},\log{T_\mathrm{eff}},\log{L}$)-space, the pre-selected sample
forms a fork-like structure, with the two hands being RC and RGB sequences. 
The $f(T_\mathrm{eff},\mathrm{[Fe/H]})$ criterion defines a plane that separates one 
sequence from another. In order to avoid the need to manually adapt  
the $f(T_\mathrm{eff},\mathrm{[Fe/H]})$ criterion each time we change the model parameters 
or isochrones, we introduced a numerical routine to 
separate the RC and RGB stars (included into the \texttt{jjmodel} code). 
After applying our routine to the pre-selected sample, 
we got a clean RC population. At the subsequent steps, we modified the vertical 
spatial densities of this complete sample to account for the geometry of the 
observational data (\mbox{Section \ref{sect:cadfs}}) and then used these vertical density
profiles for the AMR reconstruction (\mbox{Section \ref{sect:amr_fit}}). 

\subsubsection{Cumulative age distributions}\label{sect:cadfs}

To perform realistic modelling of some stellar population, it is important 
to properly take into account possible selection effects, 
so we address this question before modelling RC CADFs. 

The $R$-$z$ plot of our APOGEE samples in \mbox{Figure \ref{fig:rc_data}}
(bottom) shows that the selected RC stars are distributed in space very unevenly. 
Indeed, the APOGEE survey does not provide a full sky coverage but targets a set 
of `grid' and `non-grid' pointings, whose selection is motivated by  
location of objects of special interest and goals of the different scientific projects, 
as well as some technical limitations \citep{zasowski13,zasowski17}. 
To observe spectra, fibre-plugged plates are used, 
such that each fibre corresponds to a certain area in the sky. As a result, 
space coverage of the APOGEE consists of so-called pencil beams probing the MW stellar populations within the selected directions \citep{wilson19}. 

We can formulate this spatial selection in the following way: at each radius, 
the observed vertical number density law deviates from the real one, 
and the residual reflects the sample geometry (assuming that there are no other 
losses). In the \mbox{JJ model}, on the contrary, we by default work with 
complete populations whose vertical number density declines smoothly with $z$. 
So we need to understand whether the spatial irregularity of the APOGEE RC 
sample is relevant for the calculation of CADFs, and in which form it can be taken 
into account. 

The thick disk is known to be a well-mixed population -- quite homogeneous in age,
chemical composition, and kinematics \citep{lee11,haywood13,hayden15,hayden20}. 
Within \mbox{the JJ-model} framework, it is represented with a single 
isothermal population at each $R$. 
This implies that the predicted thick-disk age and MDs 
are the same at all heights at a given radius. Thus, even if the predicted thick-disk
vertical number density law (complete sample) does not exactly follow the observed
vertical distribution (`pencil beams'), this cannot introduce any bias to the modelled
age distribution, and therefore, no correction for the sample geometry is needed.

In the case of the thin disk, the situation is different. 
As we represent the thin disk with multiple isothermal populations, its kinematic, chemical, 
and age properties significantly change with $z$ at a fixed $R$. 
Therefore, to model the realistic age distributions of the thin-disk APOGEE RC sample, 
it is important to properly account for the sample's spatial geometry. 

To do so, we use the following approach. 
At each radius, we calculated $N_\mathrm{data}(|z|)$, 
a distribution of the observed thin-disk RC stars over $|z|$ up to \mbox{2 kpc} 
with a step of \mbox{50 pc}. 
In the model, we do not model the real star counts, but predict 
the vertical number density of RC stars according to \mbox{Eq. (\ref{eq:Nzv})}, 
$N^\mathrm{V}_\mathrm{model}(|z|)$. To be able to compare modelled and 
observed quantities, we normalised both vertical laws to their midplane values. 
As $N^\mathrm{V}_\mathrm{model}(|z|)/N^\mathrm{V}_\mathrm{model}(0) = N_\mathrm{model}(|z|)/N_\mathrm{model}(0)$, 
we arrived at the comparable quantities. The observed and predicted normalised vertical
profiles are shown at the top panel of \mbox{Figure \ref{fig:z_weights}}. 
We calculated the ratio: 
\begin{equation}
    \omega_\mathrm{z} = \left. \frac{N_\mathrm{data}(|z|)}{N_\mathrm{data}(0)} \middle/
            \frac{N_\mathrm{model}(|z|)}{N_\mathrm{model}(0)} \right..
    \label{eq:z_weights}
\end{equation}
At each $|z|$, quantity $\omega_\mathrm{z}$ shows by how much the observed vertical distribution of stars deviates from the vertical density law of a complete population. 
Assuming that $\mathrm{TM0}$ predictions are entirely correct, this deviation 
reflects the observed sample's incompleteness and geometry. Thus, we can use $\omega_\mathrm{z}$
to modify the modelled RC age distributions to mimic the geometry and incompleteness 
of the APOGEE RC sample. 

For the reasons explained below in \mbox{Section \ref{sect:amr_fit}}, we find it useful 
to test an impact of the warp on our modelling. We adopted a simple linear warp model 
from \citet{chenx19}. The normalised vertical number density laws 
and the relative weights $\omega_\mathrm{z}$ 
corresponding to $\mathrm{TM0}$ with and without the warp are shown 
in \mbox{Figure \ref{fig:z_weights}}. The mock RC CADFs calculated with $\mathrm{TM0}$ 
without the warp and weights $\omega_\mathrm{z}$ taken into account are shown in \mbox{Figure \ref{fig:rc_cadf}}. The displayed CADFs correspond to 
the 5\textsuperscript{th} iteration (consistent with the reconstructed AMR in 
\mbox{Section \ref{sect:amr_fit}}).

\subsubsection{AMR fitting process}\label{sect:amr_fit}

\begin{figure}[t]
\includegraphics[width=\columnwidth]{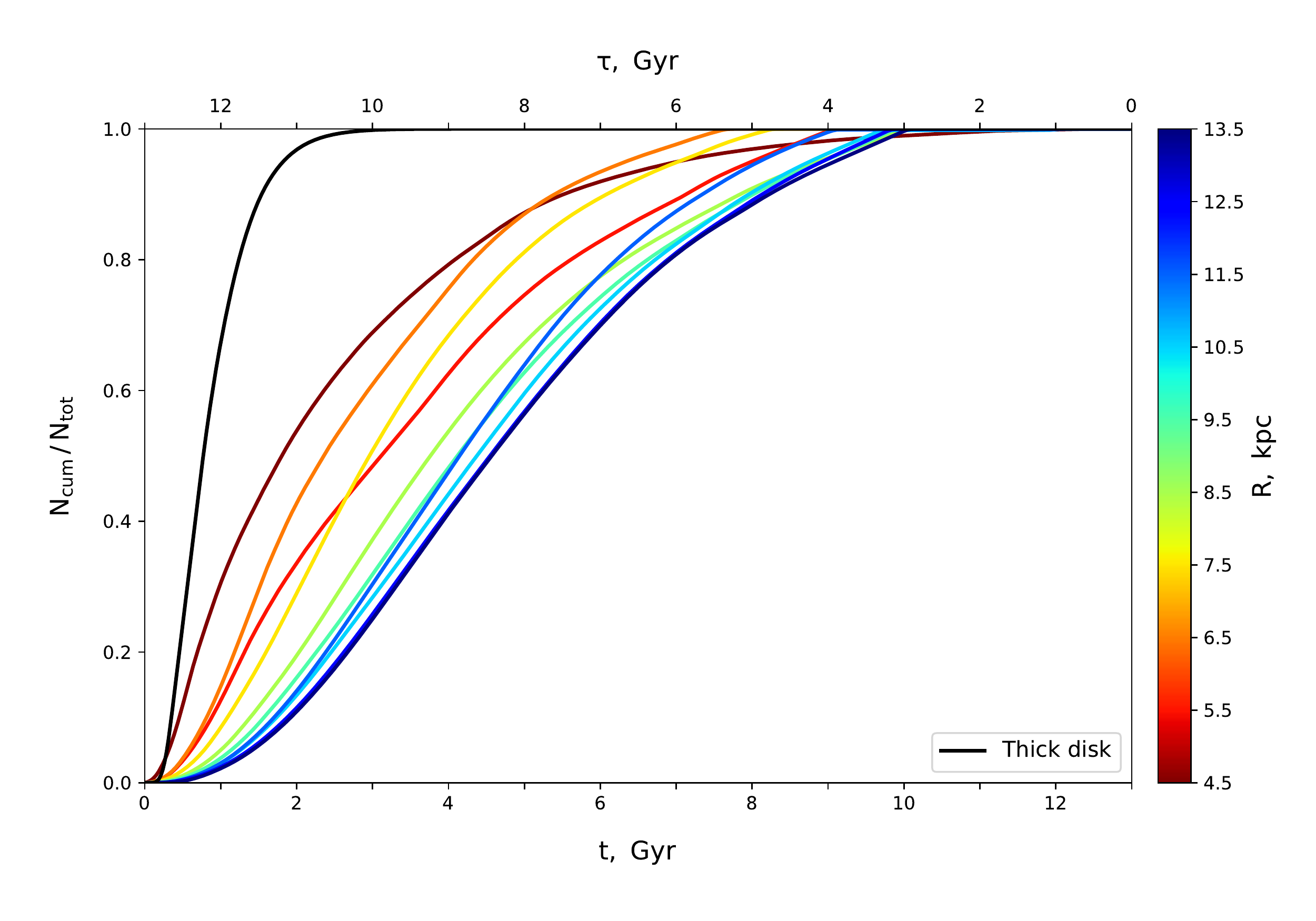}
\caption{Normalised CADFs of the thick-disk (black curve) 
and thin-disk (colour-coded) RC sample. The thin-disk radial grid corresponds to 
$\mbox{run0}$.}
\label{fig:rc_cadf}
\end{figure}

\begin{figure}[t]
\includegraphics[width=\columnwidth]{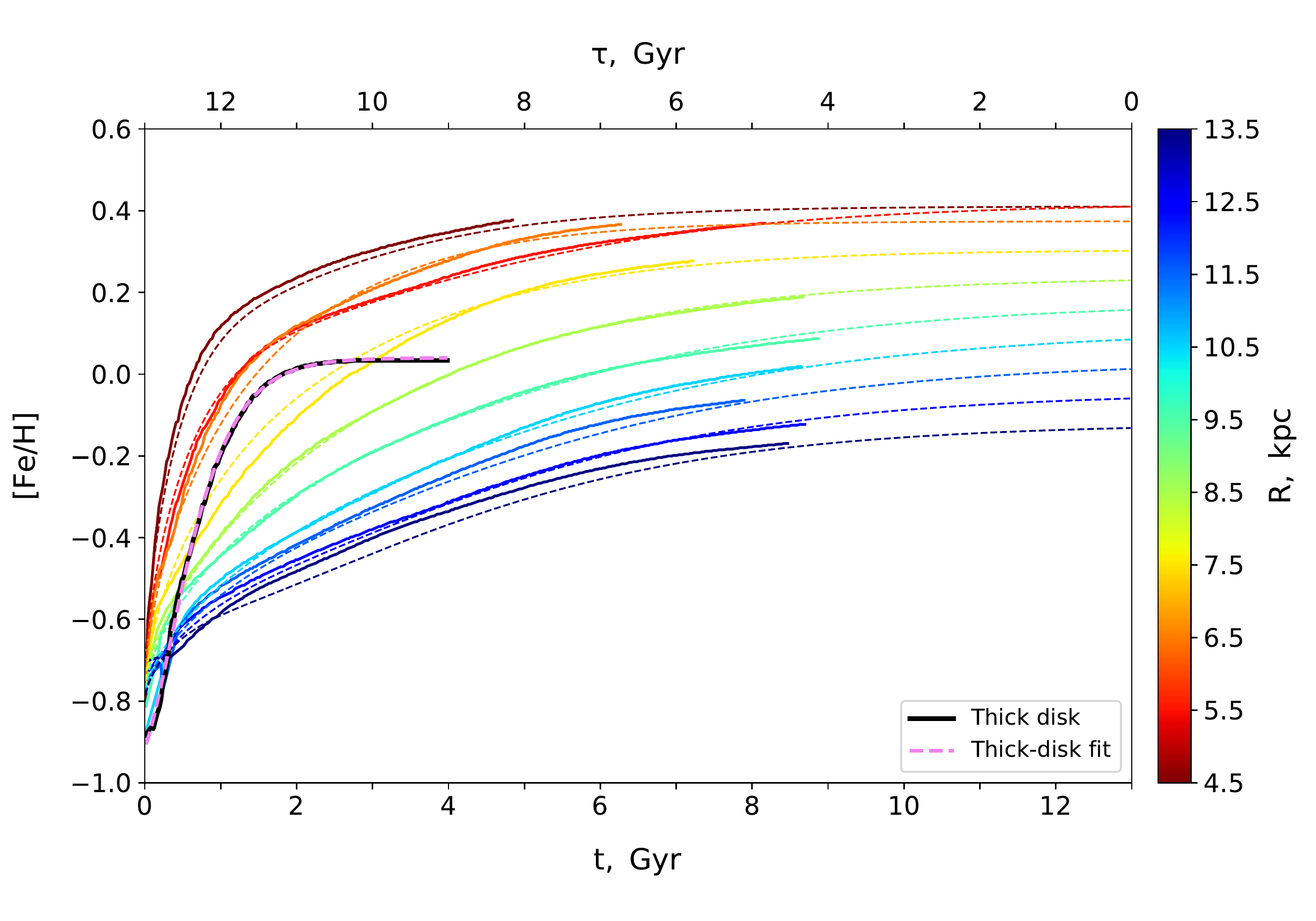}
\caption{Reconstructed AMR of the thin and thick disk (solid coloured and black lines, respectively) 
and their best fits (dashed lines) 
(\mbox{Eqs. (\ref{eq:amr_fit})-(\ref{eq:f_amrt})}, parameters from \mbox{Table \ref{tab:amr_params}}).}
\label{fig:amrdt_fits}
\end{figure}

In \citetalias{just10}--\citetalias{sysoliatina21}, we used a logarithmic function to describe 
a monotonous increase of the thin-disk sub-populations' metallicity with time, 
and in \citetalias{sysoliatina21} we proposed a $\tanh$-law for the AMR of the thick disk. 
In this study, we find that the shape of the thin-disk AMR is very different at the different $R$, 
ranging from a function with a steep increase at small $t$ in the inner disk to a quasi-linear law 
in the outer disk (\mbox{Figure \ref{fig:amrdt_fits}}). Though our old $\log$-law is 
well-suited for reproducing the local thin-disk AMR, it is unable to describe such a family of curves. 
Therefore, for the radially dependent AMR of the thin disk, we introduce a new seven-parameter 
$\tanh$-function, which can be viewed as a more complex form of the thick-disk $\tanh$-law. 
As before, the metallicity as a function of time for both thin and thick disk 
is expressed in the following form:
\begin{equation}
\mathrm{[Fe/H]}(t) = \mathrm{[Fe/H]}_0 + 
                    (\mathrm{[Fe/H]}_\mathrm{p} - \mathrm{[Fe/H]}_0) \cdot f(t),
\label{eq:amr_fit}
\end{equation}
where functions $f(t)$ for the thin and thick disk are given by 
\mbox{Eqs. (\ref{eq:f_amrd}) and (\ref{eq:f_amrt})}\footnote{In \citetalias{sysoliatina21},
we presented the thick-disk $f(t)$ without the normalisation factor as 
$t_0 \approx 1$, and thus, \mbox{$\tanh{(t_\mathrm{p}/t_0)} \to 1$} and can be dropped. 
Here we show the normalisation for the sake of consistency with 
\mbox{Eq. (\ref{eq:f_amrd})}}, respectively:
\begin{equation}
f(t) = \alpha_w \frac{\tanh^{r_{1}} \left(\sfrac{t}{t_{01}}\right)}
                       {\tanh^{r_{1}} \left(\sfrac{t_\mathrm{p}}{t_{01}}\right)} +\ 
       (1 - \alpha_w) \frac{\tanh^{r_{2}} \left(\sfrac{t}{t_{02}}\right)}
                          {\tanh^{r_{2}} \left(\sfrac{t_\mathrm{p}}{t_{02}}\right)}
\label{eq:f_amrd}
\end{equation}
\begin{equation}
    f(t) = \frac{\tanh^{r} \left( \sfrac{t}{t_0} \right)}
                {\tanh^{r} \left( \sfrac{t_\mathrm{p}}{t_0} \right).} 
\label{eq:f_amrt}
\end{equation}
Figure \ref{fig:amrdt_fits} shows the thin- and thick-disk AMR (solid curves), 
which we reconstructed following the procedure described in the previous sections. 
We note that AMR not only of the thick disk, 
whose SF ceased in the model \mbox{$\sim$9 Gyr} ago, but also of the thin disk 
does not extend to the present day. In the case of the thin disk, this reflects 
the age span of the modelled 
RC sample: depending on the radius, the last \mbox{$\sim$0.5--5 Gyr} of the thin-disk evolution are not represented by our sample (see age distributions in 
\mbox{Figure \ref{fig:rc_cadf}}). We recovered the missing ages 
by fitting the reconstructed raw AMR and extrapolating these fits to the present day. 

Fitting the thick-disk AMR with \mbox{Eqs. (\ref{eq:amr_fit}) and (\ref{eq:f_amrt})} 
is straightforward. 
The obtained parameters are listed in Table \ref{tab:amr_params}, and the fitted AMR is shown in \mbox{Figure \ref{fig:amrdt_fits}}. 

The analytic thin-disk AMR has to be a function of time $t$ and galactocentric 
distance $R$. The presented thin-disk AMR equation, \mbox{Eqs. (\ref{eq:amr_fit}) and (\ref{eq:f_amrd})}, is a function of time 
only, as it describes the AMR shape at some fixed $R$. To add dependence on the distance,
we assumed its parameters to be simple functions of $R$. 
Our tests showed that if all thin-disk AMR parameters are fitted simultaneously, 
then the corresponding generalised analytic AMR significantly deviates from the
raw AMR, which we reconstructed based on the APOGEE data.
To find the generalised AMR that is well-consistent with the raw reconstructed AMR, 
we fit parameters step by step in the following way.
% A&A version

\begin{figure*}[t]
\includegraphics[scale=0.48]{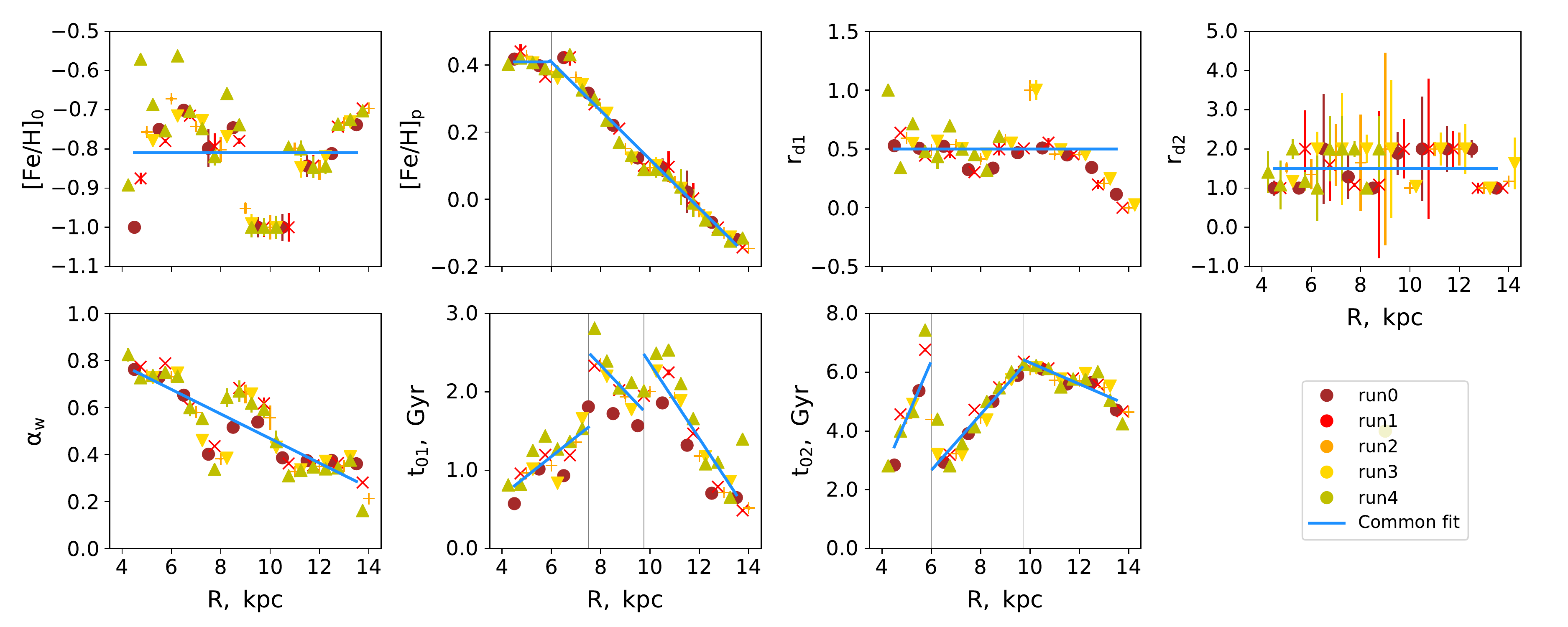}
\caption{Best-fit parameters of the thin-disk AMR. 
Different colours and symbols correspond to the five 
radial grids ($\mathrm{run0}$--$\mathrm{run4}$). Open brown circles show parameters 
for $\mathrm{run0}$ with the included warp. Vertical grey lines mark the break-point 
positions. Blue lines are the overall fits of the radial trends.}
\label{fig:amrd_params}
\end{figure*}

\ifx
Firstly, we fit the raw thin-disk AMR in each radial bin with 
\mbox{Eqs. (\ref{eq:amr_fit}) and (\ref{eq:f_amrd})}.
 
After that, we adopted analytic radial profiles for the obtained best-fit initial 
and present-day  metallicity, $\mathrm{[Fe/H]}_0$ and $\mathrm{[Fe/H]}_\mathrm{p}$. 
As one can see from $\mathrm{[Fe/H]}_0$ panel at \mbox{Figure \ref{fig:amrd_params}}, 
there is no clear radial trend and a large scatter. In order not to over-interpret 
the obtained results, we adopted a constant value for this parameter for all radii. 
$\mathrm{[Fe/H]}_\mathrm{p}$ manifests, on the contrary, a clear negative gradient 
on almost the whole interval of the examined galactocentric distances. 
In the inner disk, at $\mathrm{R} \lesssim 6$ kpc, the present-day thin-disk metallicity 
remains approximately constant. This change of $\mathrm{[Fe/H]}_\mathrm{p}$ slope can be 
related to the real physical processes, in particular, to the impact of bulge. 
Therefore, we modelled $\mathrm{[Fe/H]}_\mathrm{p}$ with a broken law, which is 
a constant  at $R < R_\mathrm{br1}$ with \mbox{$R_{br1}=6$ kpc} and linearly 
declines with radius at larger $R$. 
Then, treating $\mathrm{[Fe/H]}_0$ and $\mathrm{[Fe/H]}_\mathrm{p}$ as known, 
we again fit the raw AMR and updated the optimal values of the remaining five parameters
($r_1,r_2,\alpha_\mathrm{\omega},t_{01},t_{02}$).
 
During the fitting at the previous step, we searched for the best values 
of power indices $r_1$ and $r_2$ in the narrow non-overlapping intervals centred 
at 0.5 and 1.5, such that $\tanh$-terms in 
\mbox{Eqs. (\ref{eq:amr_fit}) and (\ref{eq:f_amrd})} dominate 
at different ages. We find that without deteriorating the fit quality, we can  
adopt for all radii $r_1=0.5$ and $r_2=1.5$.  
At this stage, we re-fit the AMR to update parameters 
($t_{01}$, $t_{02}$, $\alpha_\mathrm{\omega}$).
 
Next, we fit parameter $\alpha_\mathrm{\omega}$ with a linear profile 
and updated the remaining ($t_{01},t_{02}$).
 
Finally, we adopted fits for parameters $t_{01}$ and $t_{02}$. 
Two bottom panels at \mbox{Figure \ref{fig:amrd_params}} show their 
best-fit values at different $R$. 
It is clear that both of these parameters change with radius non-monotonously. 
At the same time, the value of metallicity predicted according to 
\mbox{Eq. (\ref{eq:amr_fit}) and (\ref{eq:f_amrd})} 
is very sensitive to $t_{01}$ and $t_{02}$. As a result, it is not possible 
to fit the obtained data points with a constant or linear trend as we did 
with $\mathrm{[Fe/H]}_\mathrm{0}$ and $\alpha_\mathrm{\omega}$ and obtain 
a good fit of the raw AMR. 
For this reason, for both $t_{01}$ and $t_{02}$ we used three-slope broken linear laws.
\fi

% Arxiv version
\begin{itemize}
\setlength\itemsep{0.25em}
    
    \item Firstly, we fit the raw thin-disk AMR in each radial bin with 
    \mbox{Eqs. (\ref{eq:amr_fit}) and (\ref{eq:f_amrd})}.
 
    \item After that, we adopted analytic radial profiles for the obtained best-fit initial 
    and present-day  metallicity, $\mathrm{[Fe/H]}_0$ and $\mathrm{[Fe/H]}_\mathrm{p}$. 
    As one can see from $\mathrm{[Fe/H]}_0$ panel at \mbox{Figure \ref{fig:amrd_params}}, 
    there is no clear radial trend and a large scatter. In order not to over-interpret 
    the obtained results, we adopted a constant value for this parameter for all radii. 
    $\mathrm{[Fe/H]}_\mathrm{p}$ manifests, on the contrary, a clear negative gradient 
    on almost the whole interval of the examined galactocentric distances. 
    In the inner disk, at $\mathrm{R} \lesssim 6$ kpc, the present-day thin-disk metallicity 
    remains approximately constant. This change of $\mathrm{[Fe/H]}_\mathrm{p}$ slope can be 
    related to the real physical processes, in particular, to the impact of bulge. 
    Therefore, we modelled $\mathrm{[Fe/H]}_\mathrm{p}$ with a broken law, which is 
    a constant  at $R < R_\mathrm{br1}$ with \mbox{$R_{br1}=6$ kpc} and linearly 
    declines with radius at larger $R$. 
    Then, treating $\mathrm{[Fe/H]}_0$ and $\mathrm{[Fe/H]}_\mathrm{p}$ as known, 
    we again fit the raw AMR and updated the optimal values of the remaining five parameters
    ($r_1,r_2,\alpha_\mathrm{\omega},t_{01},t_{02}$).
 
    \item During the fitting at the previous step, we searched for the best values 
    of power indices $r_1$ and $r_2$ in the narrow non-overlapping intervals centred 
    at 0.5 and 1.5, such that $\tanh$-terms in 
    \mbox{Eqs. (\ref{eq:amr_fit}) and (\ref{eq:f_amrd})} dominate 
    at different ages. We find that without deteriorating the fit quality, we can  
    adopt for all radii $r_1=0.5$ and $r_2=1.5$.  
    At this stage, we re-fit the AMR to update parameters 
    ($t_{01}$, $t_{02}$, $\alpha_\mathrm{\omega}$).
 
    \item Next, we fit parameter $\alpha_\mathrm{\omega}$ with a linear profile 
    and updated the remaining ($t_{01},t_{02}$).
 
    \item Finally, we adopted fits for parameters $t_{01}$ and $t_{02}$. 
    Two bottom panels at \mbox{Figure \ref{fig:amrd_params}} show their 
    best-fit values at different $R$. 
    It is clear that both of these parameters change with radius non-monotonously. 
    At the same time, the value of metallicity predicted according to 
    \mbox{Eq. (\ref{eq:amr_fit}) and (\ref{eq:f_amrd})} 
    is very sensitive to $t_{01}$ and $t_{02}$. As a result, it is not possible 
    to fit the obtained data points with a constant or linear trend as we did 
    with $\mathrm{[Fe/H]}_\mathrm{0}$ and $\alpha_\mathrm{\omega}$ and obtain 
    a good fit of the raw AMR. 
    For this reason, for both $t_{01}$ and $t_{02}$ we used three-slope broken linear laws.
\end{itemize}

\mbox{Figure \ref{fig:amrd_params}} shows the best values of five parameters of 
the thin-disk AMR, which we 
obtained using such a step-by-step fitting routine. We repeated this fitting procedure 
for our five $R$ grids. Using different binning results in a scatter in the parameters'
radial trends, but overall the data points of all runs are well consistent with 
one another; therefore the generalised AMR that we defined based on these 
profiles is not biased by the choice of bin edges. 

There is a distinct change of slope in both $t_{01}$ and $t_{02}$ trends 
around \mbox{$\sim$10 kpc}. In order to check whether this feature partially 
or fully originates from the absence of the warp in our model, we tested $\mathrm{TM0}$ 
with the warp. It is already clear from \mbox{Figure \ref{fig:z_weights}} that the impact 
of warp starts to be noticeable only at \mbox{$R \approx 13$ kpc} and is very small. 
The best-fit values corresponding to the AMR reconstruction with $\mathrm{TM0}$ with 
the warp are essentially indistinguishable from the values obtained for the runs without 
the warp at \mbox{Figure \ref{fig:amrd_params}}. 
This essentially does not change if we use the model with the more pronounced warp 
from \citet{poggio20}. From this we can at least conclude that 
the non-monotonous behaviour of $t_{01}$ and $t_{02}$ radial profiles in the outer disk 
is not related to the lack of warp in our model. 

\begin{table*}
 \centering
 %\tiny
 \caption{AMR parameters corresponding to Eqs. (\ref{eq:amr_fit})--(\ref{eq:f_amrt}). 
 Here $R^\prime=R/R_\odot$ and \mbox{$R_\mathrm{br1}$ = 6 kpc}, 
 \mbox{$R_\mathrm{br2}$ = 7.5 kpc}, \mbox{$R_\mathrm{br3}$ = 9.75 kpc}. 
 Sub-rows for $r$ and $t_0$ for the thin disk correspond to parameters 
 ($r_1$, $r_2$) and ($t_{01}$, $t_{02}$) in \mbox{Eq. (\ref{eq:f_amrd})}.}
 \begin{tabular}{c|c|c|c|c|c} \hhline{======}
   & {\begin{minipage}[c][0.6cm][c]{1.0cm} \center{$\mathrm{[Fe/H]_0}$} \end{minipage}} & 
   {\begin{minipage}[c][0.6cm][c]{3.5cm} \center{$\mathrm{[Fe/H]_p}$} \end{minipage}} & 
   {\begin{minipage}[c][0.6cm][c]{1.0cm} \center{$\alpha_\mathrm{\omega}$} \end{minipage}} & 
   {\begin{minipage}[c][0.6cm][c]{0.65cm} \center{$r$} \end{minipage}} & 
   {\begin{minipage}[c][0.6cm][c]{3.5cm} \center{$t_0$} \end{minipage}} \\ \hline
   
   \multirow{2}{*}[-1.25em]{\begin{minipage}[c][1.0cm][c]{1.5cm} Thin disk \end{minipage}} &
   \multirow{2}{*}[-1.25em]{{\begin{minipage}[c][1.0cm][c]{1.0cm} \center{$-0.81$} \end{minipage}}} & 
   \multirow{2}{*}[-1.25em]{{\begin{minipage}[c][1.0cm][c]{4.0cm}
                                $R \leq R_\mathrm{br1}: 0.41$ \\   
                                $R > R_\mathrm{br1}: -0.59 \, \mathrm{R^\prime} + 0.85$ \end{minipage}}} & 
   \multirow{2}{*}[-1.25em]{{\begin{minipage}[c][1.0cm][c]{2.3cm} 
                                \center{$-0.43 \, R^\prime + 0.99$} \end{minipage}}} & 
   {\begin{minipage}[c][0.6cm][c]{0.65cm} \center{$0.5$} \end{minipage}} & 
   {\begin{minipage}[c][1.5cm][c]{5.2cm} 
                                $R \leq R_\mathrm{br2}: 2.04 \, R^\prime - 0.32$ \\   
                                $ R_\mathrm{br2} < R \leq R_\mathrm{br3}: -2.71 \, R^\prime + 4.99$ \\
                                $R > R_\mathrm{br3}: -3.91 \, R^\prime + 7.14$ \end{minipage}}  \\ \cline{5-6}

   & & & & {\begin{minipage}[c][0.6cm][c]{0.6cm} \center{$1.5$} \end{minipage}} & 
   {\begin{minipage}[c][1.5cm][c]{5.2cm} 
                                $R \leq R_\mathrm{br1}: 15.89 \, R^\prime - 5.29$ \\   
                                $ R_\mathrm{br1} < R \leq R_\mathrm{br3}: 7.77 \, R^\prime - 3.04$ \\
                                $R > R_\mathrm{br3}: -2.97 \, R^\prime + 9.96$ \end{minipage}} \\ \hline
   
   {\begin{minipage}[c][0.6cm][c]{1.5cm} Thick disk \end{minipage}} & 
   {\begin{minipage}[c][0.6cm][c]{1.0cm} \center{$-0.91$} \end{minipage}} & 
   {\begin{minipage}[c][0.6cm][c]{1.0cm} \center{$0.04$} \end{minipage}} & -- & 
   {\begin{minipage}[c][0.6cm][c]{0.65cm} \center{$1.28$} \end{minipage}} & 
   {\begin{minipage}[c][0.6cm][c]{1.0cm} \center{$0.9$} \end{minipage}} \\ \hhline{======}
 
 \end{tabular}
 \label{tab:amr_params}
\end{table*}

As a wrap-over, we applied the step-by-step fitting to the parameter values of 
$\mathrm{run0}$--$\mathrm{run4}$ collectively and obtained the final radial profiles 
of the AMR parameters (green lines on \mbox{Figure \ref{fig:amrd_params}}). 
At this stage, we also optimised the break-point positions for $t_{01}$ and $t_{02}$ 
by minimising a mean relative deviation between the real and fitted best parameter values:
\begin{equation}
\frac{\langle |t_{01} - f_{t_{01}}| \rangle}{\langle \sigma_{t_{01}} \rangle} + \ 
\frac{\langle |t_{02} - f_{t_{02}}| \rangle}{\langle \sigma_{t_{02}} \rangle} \to \mathrm{min} \nonumber
 \label{eq:rbr_min}
\end{equation}
Here $f_{t_{0i}}$ are the analytic profiles with some assumed break-point positions,
$\sigma_{t_{0i}}$ is the dispersion. The calculation was performed in the $R$ grid 
with bin centres at $4.5,5.5,...,14.5$ kpc. 
We find that the optimal position of the second break is \mbox{9.75 kpc} for both parameters, 
while the first break lies at \mbox{7.5 kpc} and \mbox{6.25 kpc} for 
$t_{01}$ and $t_{02}$, respectively. 
The break point at \mbox{6.25 kpc} is very close to \mbox{$R_{br1} = 6$ kpc}, which 
we adopted for $\mathrm{[Fe/H]}_\mathrm{p}$, as this parameter is strongly 
correlated with $t_{02}$ (both affect the AMR shape at young ages). 
In order not to further complexify our analytic representation of the thin-disk AMR, 
we set the first break of $t_{02}$ profile also to \mbox{6.0 kpc}. 

The obtained fits for the AMR parameters are listed in \mbox{Table \ref{tab:amr_params}}.
The resulting generalised thin-disk AMR is plotted in \mbox{Figure \ref{fig:amrdt_fits}}
with dashed coloured curves. We see that the final AMR fit has a very good consistency 
with the raw reconstructed AMR.

\subsubsection{Consistency test}\label{sect:consistency}

\begin{figure*}[t]
\includegraphics[scale=0.55]{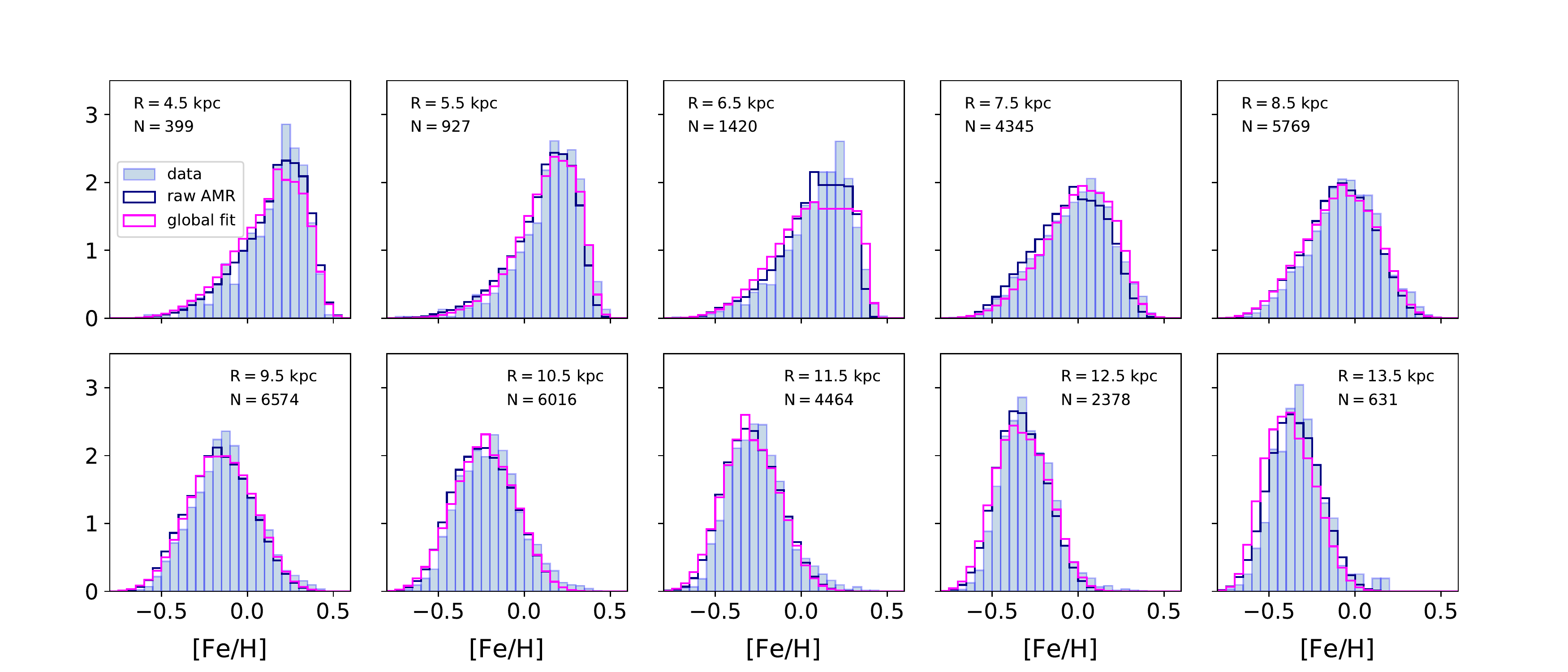}
\caption{Normalised observed and predicted MDs of the thin-disk RC sample 
calculated for the different radial bins. 
Dark blue and magenta step histograms are the \mbox{JJ-model} predictions 
that correspond to the raw reconstructed AMR and the generalised analytic AMR
calculated according to \mbox{Eqs. (\ref{eq:amr_fit}) and (\ref{eq:f_amrd})} with 
the best-fit parameters from \mbox{Table \ref{tab:amr_params}}.}
\label{fig:md_thin}
\end{figure*}

\begin{figure}[t]
\includegraphics[width=\columnwidth]{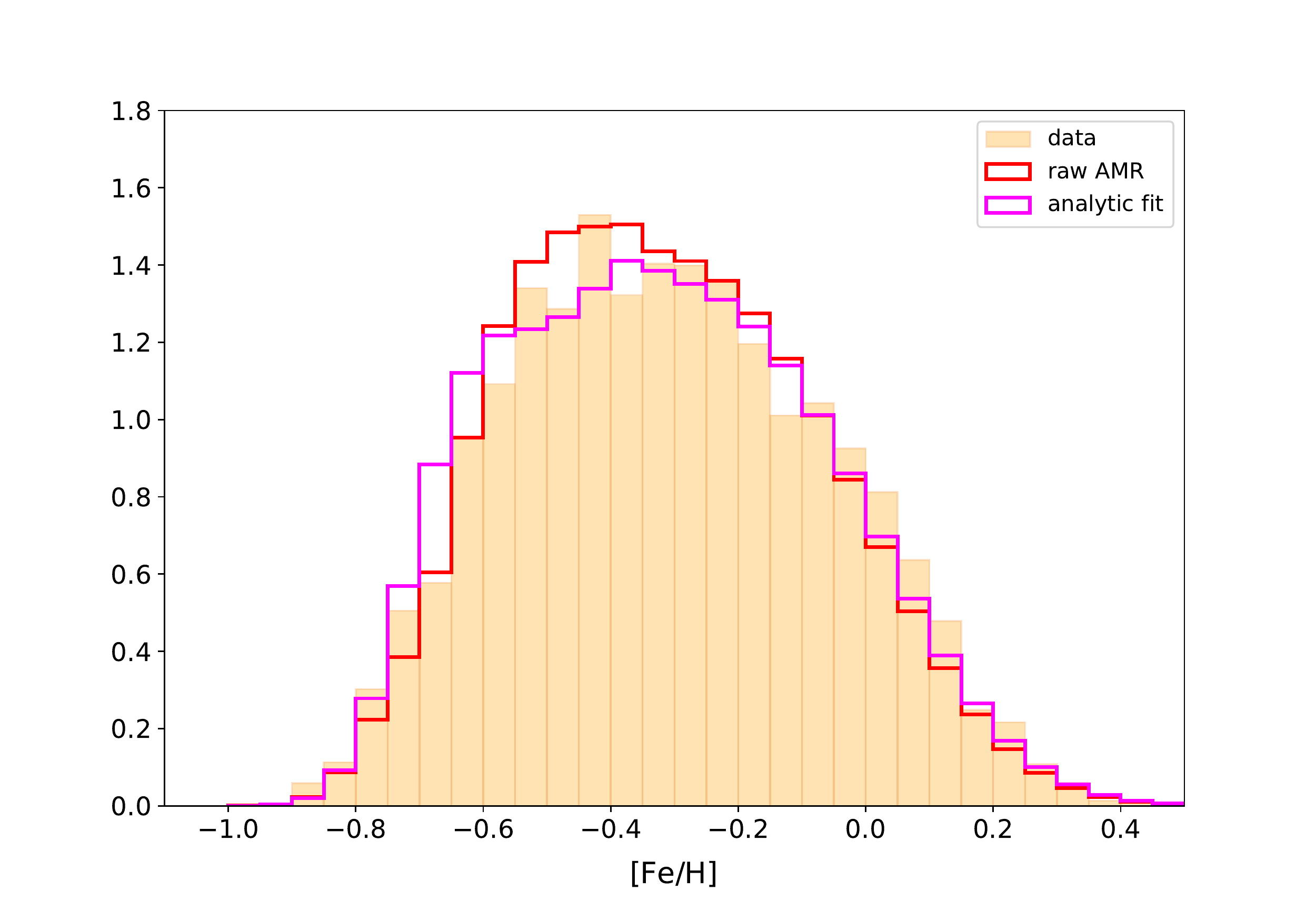}
\caption{Same as in \mbox{Figure \ref{fig:md_thin}}, but for the thick disk. 
Both observed and predicted MDs correspond to the range of galactocentric distances 
\mbox{4--14 kpc}.}
\label{fig:md_thick}
\end{figure}

To demonstrate the consistency between the APOGEE data and \mbox{the JJ model} after the AMR calibration, 
we predicted the RC MDs using the derived AMR and compared them 
to the observed metallicities. 
The MD modelling was performed in the full consistency with the approach 
used to reconstruct the AMR function. For the thick disk, MD was predicted for 
the mock RC population within the range of \mbox{4--14 kpc} and at \mbox{$|z|<2$ kpc} 
in the $\mathrm{TM0}$ framework. For the thin disk, MDs were calculated separately in 
each radial bin, and the vertical densities of the modelled RC populations 
were weighted with $\omega_\mathrm{z}$. We also added an error model as a post-processing
step: each MD was converted into cumulative form, convolved with a Gaussian kernel 
of $\sigma_\mathrm{[Fe/H]}$ as derived from the data (\mbox{Figure \ref{fig:rc_cmdf}}),
and then transformed back to non-cumulative distribution. 
The resulting thin- and thick-disk MDs are shown in 
\mbox{Figures \ref{fig:md_thin} and \ref{fig:md_thick}}. Two types of step histograms
correspond to $\mathrm{TM0}$ predictions with the raw and generalised analytic AMR. 

We see that the model and data are well consistent in terms of the overall shapes and 
peak positions of the distributions. The maximum shift that we report between 
the modelled and observed MD is only \mbox{$\sim0.05$ dex}.  

\section{Discussion}\label{sect:discussion}

The goal of the JJ model is to provide a flexible modelling tool for 
the main body of the MW disk in terms of stellar populations and 
their properties. We model the axisymmetric disk
in dynamical equilibrium. In order to allow the investigation of the Galaxy 
evolution over the full age range, we represent the stellar content of the disk 
by sequences of stellar sub-populations with a high time resolution of 
\mbox{25\,Myr}. This ansatz is complementary to \mbox{the BGM model}, 
which tries to incorporate all MW features including the bar, spiral structure, 
and the warp at the cost of a restricted time resolution. The input \mbox{JJ-model}
functions SFR, AVR, and AMR are relatively simple functions of galactocentric
distance and time (i.e. age). The IMF is assumed to be universal, but this could 
easily be made a function of metallicity in the framework of \mbox{the JJ model} 
(see \mbox{Section \ref{sect:thindisk_imf}}). 

\subsection{Model properties and limitations}\label{sect:discussion_model}

The JJ model provides predictions for the density distribution, vertical velocity
distribution function, age distributions, and MD function 
for stellar subsamples (along the MS, RC stars, etc.). The parameters of the input
functions are constrained by comparison of these predictions with corresponding
observed samples. We did not use observed samples with individual age determinations,
because they are still very restricted and may have significant biases. Therefore,
direct age determinations can be used as independent tests of the model 
(or our model can be used as a tool for deriving age distributions of stellar populations).
We only remark here that to perform a robust model-to-data comparison, 
the impact of dust must be taken into account: a three-dimensional 
extinction map can be used to redden 
our synthetic photometry to compare to the data (forward modelling, 
\citealp{sysoliatina18}), or the data can be preliminary de-reddened and then compared to the model 
(as in \citetalias{sysoliatina21}). Also, the selection function of the sample needs to be 
thoroughly scrutinised to include the corresponding selection effects to the model.

Since the JJ model contains only axisymmetric populations in dynamical equilibrium, 
it can be used to characterise the properties of model-to-data deviations, such as 
the spiral arms and the warp. Also, due to the modular design of the code, 
simple spiral arm or warp models can be implemented. As an example, we tested 
the impact of the warp on the AMR derivation and found no significant differences 
(see \mbox{Section \ref{sect:amr_fit}}).

The SFR, AVR, and AMR describe the present-day properties of stellar sub-populations
at the current galactocentric distance. All three functions depend on each other 
for the derivation of the present-day distribution functions to be consistent with
observations. Also the stellar birth places do not appear explicitly in our model.
Only in the limiting case of negligible radial migration (blurring and churning), 
the time- and radially dependent SFR is the star formation history at radius $R$, 
the AVR represents the heating function, and the AMR corresponds to the metal
enrichment law. In the case of a significant radial migration impact, these input
functions can be viewed as the results of a convolution of the true SFR, AVR, and 
AMR with the radial migration model. In \citetalias{sysoliatina21}, we 
addressed this topic when discussing physical meaning of the reconstructed local SFR
of the thin disk. Assuming that the impact of migration is a function of time, 
we can formulate this the other way: specify a range of radii that are represented 
by a particular age interval of the SFR, AVR, or AMR. 

Additionally, the JJ model can be combined with the \textit{Chempy} code to derive 
the chemical enrichment history at each radius including the gas infall 
and abundances of other elements.

\subsection{AMR and radial migration} \label{sect:discussion_amr}

Abundance distributions of different elements play a fundamental role in determining
the evolution of galaxies. Here we used only the iron abundance and the 
$\alpha$ enhancement in a simple way to separate the thin and thick disk and derive
the AMRs. Many other elements are available to test the local enrichment model 
without radial migration. If radial migration is relevant, then the interpretation 
of the AMR and the other input functions is more complex as we explained above. 
The unique AMR, which we derived, corresponds to the mean metallicity 
as a function of age, and the underlying real AMR is a two-dimensional distribution 
as derived in many publications based on individual stellar ages 
(e.g. \citealp{bergemann14}). Our findings for the AMR are in general comparable 
to the picture presented in \citet{feuillet19} where the disk AMR was reconstructed
at different $R$ and $z$ based on the \textit{Gaia} DR2 and APOGEE data. 
They also see a significant variation of the AMR across the disk with the 
outer disk being more metal-poor. Their spatial age distribution gives evidence 
for disk flaring, which we also reproduce in our model. 

Local mixing (blurring) produced by the non-circularity of the orbits can be taken
into account by replacing the current radius by the guiding radius of the orbit. 
For the population properties at a given current radius a distribution of guiding
radii has to be derived. In a similar way, radial migration working over larger
distances via resonances (churning) can be modelled by replacing the guiding radii 
by distributions of birth radii, which define the efficiency of the migration process.
In both cases, the main challenge is to take into account the correlations between
different quantities due to the radial gradients in ages, densities, and
metallicities. 

In order to explain the large spread of abundances in the 
\mbox{[$\mathrm{\alpha/Fe}$]-[$\mathrm{Fe/H}$] plane}, 
local enrichment models can be combined with
radial migration \citep{schoenrich09,kubryk15,hayden15,johnson21}. 
\citet{minchev18} used N-body simulation model for a MW-like galaxy
extended by a local chemical enrichment model to derive the evolution and enrichment
history of the stellar disk based on a simple radial migration model. Recently, 
\citet{xiang22} investigated the early MW evolution based on the ages and abundances
of about \mbox{250\,000 subgiant} stars. They find a well-mixed 
\mbox{$\alpha$-enhanced} thick disk with a similar early formation and enrichment 
as used in our model, but with an even more extended formation history 
of \mbox{4\,Gyr} and a higher maximum metallicity. Our thin-disk model corresponds 
to their dominating \mbox{low-$\alpha$} evolution sequence. For a more detailed
comparison the spatial distribution as a function of galactocentric radii needs 
to be analysed.

\section{Conclusions}\label{sect:conclusions}

We have presented a publicly available update of our semi-analytic MW disk model, 
which is based on an iterative solving of the Poisson equation and recovers 
a self-consistent potential-density pair independently at each radius\footnote{Github repository of the \texttt{jjmodel} code: \url{https://github.com/askenja/jjmodel}}. 
We generalised the local \mbox{JJ model} previously calibrated against the 
\textit{Gaia} DR2 sample (\citetalias{sysoliatina21}) 
by introducing radial variation to such input functions 
as the thin-disk SFR, AVR, and AMR. 
Our test model, which implies inside-out disk growth, is able to mimic known trends in the 
MW disk structure, such as broken exponential radial profiles and flaring of the mono-age 
sub-populations.

As a first step towards the global calibration of \mbox{the JJ model}, 
we reconstructed the AMR using the extended APOGEE RC sample. We find that 
the shape of the thin-disk AMR changes a lot within the considered range 
of galactocentric distances, \mbox{4--14 kpc}, so we introduced a new analytic 
function that can reproduce this variation, a seven-parameter double-$\tanh$ law. 

Applicable to a large volume, \mbox{4 kpc $ \lesssim R \lesssim$ 14 kpc} 
and \mbox{$|z| \lesssim$ 2 kpc}, the model can test different SFR shapes, 
disk flaring, IMFs, and AMRs. 
The JJ model is characterised by a fine time resolution of \mbox{25 Myr} which can also be 
combined with a fine spatial resolution within the allowed range of $R$ and $z$.
Thus, the JJ model complemented by the PARSEC, MIST, 
and BaSTI isochrones can be used for recovering disk evolution, creating MW mock
catalogues, or revealing the spiral structure and any other density deviations 
from the smooth axisymmetric disk.

\section*{Acknowledgements}

{\tiny{This work was supported by the Deutsche Forschungsgemeinschaft 
(DFG, German Research Foundation)  -- Project-ID 138713538 --
SFB 881 (`The Milky Way System', subproject A06). 

%Besides the software mentioned in the text,
This research made use of Python packages SciPy \citep{scipy-nmeth20} and NumPy \citep{numpy11}, 
a Python library for publication quality graphics 
matplotlib \citep{hunter07}, a community-developed core Python package for Astronomy Astropy \citep{astropy13,astropy18}, 
a Python mini-package \texttt{fast-histogram} (\url{https://github.com/astrofrog/fast-histogram}), as well as
an interactive graphical viewer and editor for tabular data TOPCAT \citep{taylor05}. 

We also thank Akash Vani, Oleksiy Golubov, and our anonymous referee for their comments 
which helped to improve scientific quality and clarity of the paper. 

}} 

%%%%%%%%%%%%%%%%%%%%%%%%%%%%%%%%%%%%%%%%%%%%%%%%%%%%%%%%%%%%%%%%%%%%%
%%%%%%%%%%%%%%%%%%%%%%%%%%%%%%%%%%%%%%%%%%%%%%%%%%%%%%%%%%%%%%%%%%%%%
\bibliographystyle{./aa.bst}
\bibliography{jj_extended.bib}

\end{document}